\documentclass[aps,prb,twocolumn, superscriptaddress]{revtex4-1}

\usepackage{CJK}
\usepackage{graphicx}
\usepackage{amsmath,amssymb,amsfonts}
\usepackage{placeins}
\usepackage{soul}
\usepackage{color}
\usepackage{wasysym}

\begin{document}



\title{Charge transport in InAs nanowire Josephson junctions}

\author{Simon Abay}
\email{abay@chalmers.se}
\affiliation{Department of Microtechnology and Nanoscience (MC2), Chalmers University of Technology, SE-412 96 G\"oteborg, Sweden}

\author{Daniel Persson}
\affiliation{Department of Microtechnology and Nanoscience (MC2), Chalmers University of Technology, SE-412 96 G\"oteborg, Sweden}

\author{Henrik Nilsson}
\affiliation{Division of Solid State Physics, Lund University, Box 118, S-221 00 Lund, Sweden}
\author{Fan Wu}
\affiliation{Department of Microtechnology and Nanoscience (MC2), Chalmers University of Technology, SE-412 96 G\"oteborg, Sweden}
\author{H.Q. Xu}
\affiliation{Division of Solid State Physics, Lund University, Box 118, S-221 00 Lund, Sweden}
\affiliation{Key Laboratory for the Physics and Chemistry of Nanodevices and Department of Electronics, Peking University, Beijing 100871, China}
\author{Mikael Fogelstr\"om}
\affiliation{Department of Microtechnology and Nanoscience (MC2), Chalmers University of Technology, SE-412 96 G\"oteborg, Sweden}
\author{Vitaly Shumeiko}
\affiliation{Department of Microtechnology and Nanoscience (MC2), Chalmers University of Technology, SE-412 96 G\"oteborg, Sweden}
\author{Per Delsing}
\email{per.delsing@chalmers.se}
\affiliation{Department of Microtechnology and Nanoscience (MC2), Chalmers University of Technology, SE-412 96 G\"oteborg, Sweden}

\date{\today}

\begin{abstract}
We present an extensive experimental and theoretical study of the proximity effect in InAs nanowires connected to superconducting electrodes. 
We  fabricated and investigated devices with suspended gate controlled nanowires and non-suspended nanowires, with a broad range of lengths and normal state
resistances. We analyze the main features of the current-voltage characteristics: the Josephson current, excess current, and subgap current as functions of length,
temperature, magnetic field and gate voltage, and compare them with theory. The Josephson critical current for a short length device, $L=30$\,nm, exhibits a record
high magnitude of $800$\,nA at low temperature that comes close to the theoretically expected value.
The critical current in all other devices is typically reduced compared to the theoretical values. 
The excess current  is consistent with the normal resistance data and agrees well with the theory.  The subgap current shows large number of structures, some of
them are identified as subharmonic gap structures generated by Multiple Andreev Reflection. The other structures, detected in both suspended and non-suspended
devices, have the form of voltage steps at voltages that are independent of either superconducting gap or length of the wire. By varying the gate voltage in suspended 
devices we are able to observe a cross over from typical tunneling transport at large negative gate voltage, with suppressed subgap current and negative excess
current, to pronounced proximity junction behavior at large positive gate voltage, with enhanced Josephson current and subgap conductance as well as a large positive 
excess current.  
\end{abstract}

\maketitle



\section{Introduction}

Semiconducting nanowires (NW) have  been a focus of intensive research for their potential applications as building blocks in nano-scale devices.
\cite{samuelsson,clas, Lieber, Shadi} The nano-scale dimension of the semiconducting nanowires, comparable to the electronic  Fermi wave length, also makes
them  an attractive  platform for  studying the fundamental phenomena of quantum transport. By tuning the Fermi wavelength by means of electrostatic gates 
one gets access to such quantum phenomena as conductance quantization, \cite{Kouwenhoven, simon} and quantum interference effects.\cite{Jespersen}

Another research interest has been the proximity effect in nanowires induced  by connecting them to superconducting 
electrodes (S).\cite{Tinkham, Kewo} In such devices, S-NW-S, the nanowire serves as a weak link through which a supercurrent can flow due to the 
presence of the phase difference between the superconducting condensates.\cite{Kewo, Xu, Abay} 

Among a variety of nanowires tested in experiments, nanowires of InAs play a central role.\cite{Tinkham, Xu} This is due to their  material 
properties: high electron mobility, low effective mass, and pinning of the Fermi level in the conduction band that permits highly transparent galvanic 
S-NW contacts.
Hybrid devices of InAs nanowires have demonstrated Andreev subgap conductance,\cite{Takahiro, Doh, simon} Josephson field effect,\cite{Kewo, Xu}
and Cooper-pair beam splitting.\cite{Schoenenberger} More recently, the nanowire hybrid devices attracted new attention following theoretical predictions of 
Majorana bound states in NW-S proximity structures.\cite{LeoMaj, LundMaj, Anionic}

In spite of intensive research, no systematic investigation of the proximity effect in InAs nanowires has been reported, leaving open important questions 
about consistency of the observed transport phenomena and theoretical views of the proximity effect.

In this paper, we report on extensive experimental studies of current-voltage characteristics (IVC) of a large variety of hybrid devices made with InAs 
nanowires connected to aluminum electrodes. These Al-InAs NW-Al devices include suspended and non-suspended nanowires, nanowires with different 
lengths, and tested at different temperatures, magnetic fields, and gate voltages. We measured the main proximity effect characteristics: Josephson critical current,
excess current,  and subgap current features,  and we make a quantitative comparison with  relevant theoretical models.  

Our main conclusion is that the most properties of the proximity effect can be qualitatively understood and quantitatively reasonably well fit on the basis of 
existing theory.  In particular, the record high Josephson critical current of 800\,nA, observed in the shortest studied nanowire (with 30\,nm separation
between superconducting electrodes)  is close to the theoretical bound for ballistic point contacts.\cite{Kulikl}  
In longer devices, the decay of the critical current with length is consistent with a
cross over from ballistic to diffusive transport regime, followed by cross over from a short- to long-junction behavior. 

In the gate controlled suspended devices
we observe a cross over from a distinct S-normal metal-S (SNS) type behavior with large positive excess current and enhanced subgap conductance to tunneling
S-inslulator-S (SIS) type behaviour in accordance with gradual depletion of the conducting channels by the gate potential and increase of the wire resistance. 

We also observe subgap current features associated with Multiple Andreev Reflection (MAR) transport.\cite{Klapwijk1982} In addition to those MAR features we 
systematically observe subgap features, which are not associated with MAR but have some different origin. These features are not related 
to phonon-induced resonances,\cite{Andrey} and they do not seem to have an electromagnetic origin. They appear on the IVCs as voltage steps, strikingly 
similar to the voltage steps generated by phase slip centers in superconducting whiskers.\cite{Meyer,Skocpol}   

The structure of the paper is as follows: After describing the device fabrication and experimental setup in Section II, we summarize  
the normal state conduction properties of the devices in Section III, which give the necessary input for choosing an appropriate theoretical model of the proximity effect
described in Section IV. Then in the following sections we discuss the superconducting transport properties: excess current in Section V, Josephson current  
in Section VI, sub gap current, and variation of the current under the gate potential in Section VII. Section IX contains conclusive remarks.

\section{Experimental details}

\subsection{Sample Fabrication}

\begin{figure}[ht!]
\center
\includegraphics[width=0.75\linewidth]{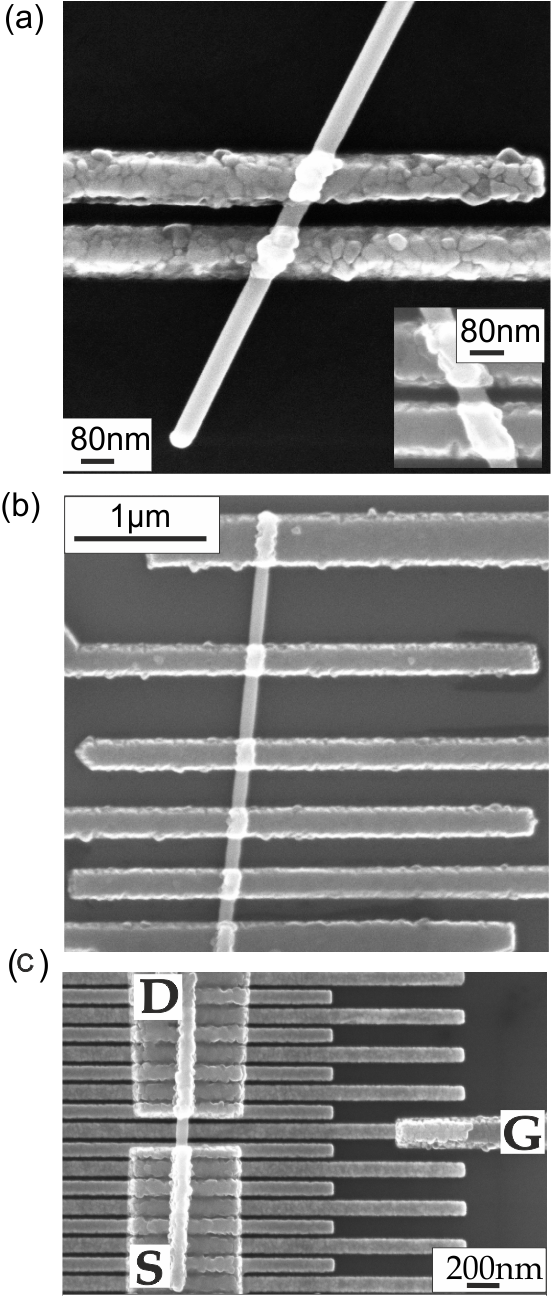}
\caption{SEM images of investigated devices.
(a) Short length device of type A defined by a single-step electron-beam lithography (inset shows a short length device defined by a double-step e-beam lithography). (b)  Device of type B has multiple contacts on a single nanowire separated by different lengths. (c) Three-terminal device with a suspended nanowire and a nearby local gate, the gate is 15\,nm below the nanowire.}
\label{fig:SEM}
\end{figure} 

The devices we have investigated are of three types: nanowires placed directly on the substrate with either a) two superconducting contacts or b) 
multiple contacts, and  c) suspended devices with local gates (Fig.\,\ref{fig:SEM}). All devices are made on standard Si substrates capped by 400\,nm 
thick SiO$_{2}$.

The nanowires are grown by chemical beam epitaxy.\cite{LundFab} In the growth process, metal-organic gaseous sources are thermally cracked to their
components and the growth materials are directed as a beam towards an InAs substrate placed in the growth chamber. At the optimal temperature, the 
nanowire growth is catalyzed by Au aerosol particles that have been distributed on the substrate. The sizes of the Au-seeds determine the diameter of 
the nanowires. In this paper, the nanowires are taken from a single growth batch with an average diameter of $80$\,nm.

To fabricate the non-suspended devices, InAs nanowires are first transferred to a Si substrate and their relative positions with respect to predefined marks 
are determined with the help of scanning electron microscope (SEM) images. The extracted locations are then used to pattern superconducting 
Ti/Al (5/150\,nm thick) contacts on top of the nanowires. Depending on the intended device length, \emph{i.e.} distance between source and drain electrodes, the
superconducting contacts are defined by either single-step or double-step electron beam (e-beam) lithography.\cite{Abay} The shorter devices ($L < 100$\,nm)
are defined by the double-step e-beam lithography whereas the longer devices ($L\geq 100$\,nm) are defined by the single-step e-beam lithography.  A SEM
image of a typical two terminal device ($L\approx$ 100\,nm defined by the single e-beam lithography) is shown in Fig.\,\ref{fig:SEM}a. The inset image shows a
short length device of $L\approx$ 60\,nm defined by the double-step e-beam lithography.

To fabricate the suspended devices, a standard Si substrate is first patterned with interdigitated Ti/Au stripes.\cite{Henrik} InAs nanowires are then
transferred to the already patterned Si substrate and some of the nanowires end up on top of the interdigitated metal stripes. The stripes are patterned in a
two-step fabrication process in order to get a height difference of 15\,nm between every two adjacent stripes. This allows the nanowires to rest on the thicker
electrodes (65\,nm thick)  while being suspended above the substrate and the thinner electrodes (50\,nm thick). With the help of SEM images, the positions of
suitable nanowires are found and superconducting electrodes Ti/Al (5/150\,nm thick) are defined on selected nanowires with e-beam lithography. A SEM image of
a suspended device is shown in Fig.\,\ref{fig:SEM}c. 

To get good transparency of the metal-nanowire interfaces, an ammonium polysulfide solution (NH$_{4}$S$_{\rm x}$) cleaning process \cite{Lund,Abay} has
been used prior to evaporation of the superconducting contacts. The samples are then characterized at room temperature and stored in a vacuum box before
further measurements at low temperatures.

\subsection{Experimental setup}

Current-voltage characteristics  of the devices are measured in a dilution refrigerator with a base temperature of 15\,mK. The IVCs are recorded in
either current or voltage bias configuration. In the current-bias mode, the current is determined by a high resistance bias resistor in series with the device. 
As we increase the current, the voltage across the device is simultaneously measured with a differential amplifier. In the voltage-bias mode, a voltage is 
directly applied across the device while the current is measured simultaneously by a transimpedance amplifier. To decrease noise coupling to the devices, 
the electrical lines in the measurement set up are well filtered and thermally anchored at different temperature stages of the refrigerator. The 
measurement setup is also designed to measure IVCs as a function of temperature and magnetic field.

\begin{figure}[hb]
\includegraphics[width=0.75\linewidth]{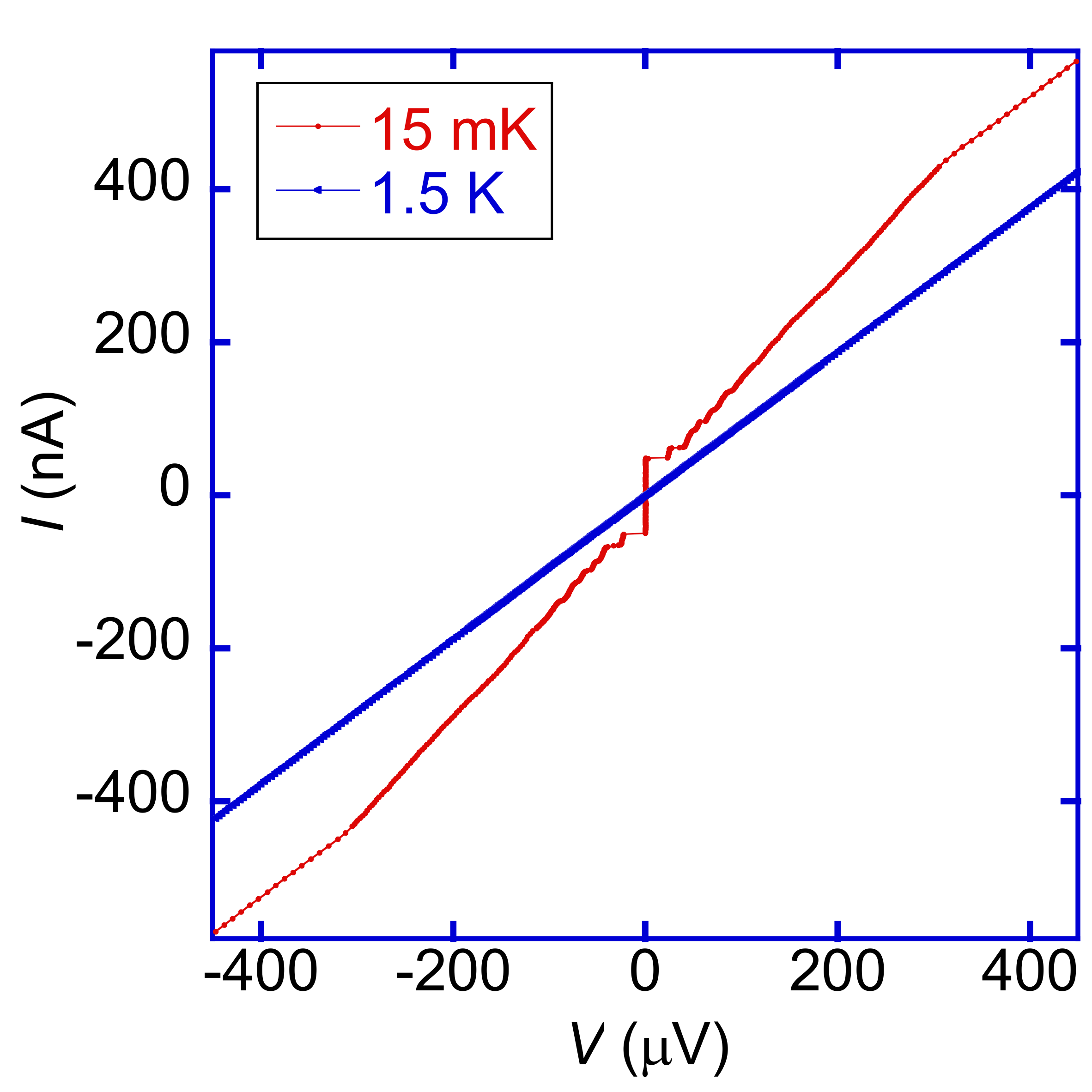}
\caption{Normal state IVC (blue line) and superconducting state IVC (red line)  for a device of length $L\approx150\,$nm (sample B$_{\rm 5a}$). 
The superconducting IVC shows normal state Ohmic behavior ($ R _{n}=1.07$\,k$\Omega$) at  $ |V| > 2\Delta/e$, enhanced conductance 
and current features at subgap voltages  $ |V|\le 2\Delta/e$,  and the Josephson current. }
\label{fig:IV}
\end{figure} 

\begin{table}[ht]
\begin{tabular}{l*{8}{c}r}
Device &  &  L(nm)                & R$_n$(k$\Omega)$ &      I$_m$(nA) &     $eI_mR_n/\Delta$  & $eI_{exc}R_n/\Delta$   \\
\hline

A               &1    &30         & 0.16          &800	         &1.02	    & 1.52  \\
                 &2    &90         & 0.55          &	95                & 0.40     &0.87  \\
                 &3    &100       &  0.56          & 54               &0.23      & 0.75 \\
                 &4    &220       &1.04	         &30	 	 &0.24    &0.76  \\
\hline
        B   &5a    &150        & 1.07		 & 50	 	 & 0.41           & 1.20  \\
                     &5b    &170        & 1.28		 &40	    	 &0.39	 &1.28 \\ 
                    &5c      &180        &1.34	 	&36	  	  &0.37 	  &1.31\\
                    &5d      &190        &1.37	           &35	   	   &0.36	  &1.11  \\

           &6a         & 110	  &1.84	 &23	 	  &0.32	 &1.21\\
          &6b         & 200	 &2.40            &12		 &0.21	 &0.81\\
           &6c        & 250	 &2.72	  &9	          	 &0.20	 &0.77\\
          &6d         & 500	&4.21	            &3		& 0.10	 &0.71\\  
           &6e        & 600	 &4.82	 &1		 &0.04	&0.64\\

\hline
C         &7 & 200       & 2.23    &15	    &0.24	   &1.30\\
           &8 & 150       & 3.3      &13	    &0.34	   &1.17\\
           &9 & 130        & 2.19   &28               &0.47	   &1.02\\
           &10& 300      & 3.80    &6                 &0.17	    &1.23\\
           &11&150	 &5.01    &2.6	     &0.10	   &0.78\\
           &12&200	 &3.6       &7.5	     &0.21	    &1.11\\

\hline

\end{tabular}
\caption{Measurement values for devices of different types.  Type A devices are nanowires with two superconducting contacts. 
Type B devices are defined on a single nanowire with multiple contacts. Adjacent junctions in the B-type devices are marked with alphabetic letters. 
Type C devices are suspended devices with local gates. The nanowires are taken from the same growth batch of approximately 80\,nm in diameter.}
\end{table}

\subsection{Current-voltage characteristics}

A typical IVC is shown in Fig.\,\ref{fig:IV}  for a device of  length $L\approx150\,$nm (sample B$_{5a}$ in Table I). Above the 
critical temperature the IVC (blue) exhibits Ohmic behavior with a normal-state resistance of $ R _{n}=1.07$\,k$ \Omega$.
The critical temperature, $T_c=1.1$\,K,  for the devices was determined from samples with shorted electrodes, {\it i.e.}
without any nanowire. This value agrees well with the temperature at which the Josephson current disappears in the samples 
with strong Josephson coupling. At temperatures well below
$T_{c}$, the IVC (red)  shows three distinct conductance regimes. i)  For voltages $ |V| > 2\Delta/e$, the IVC shows a linear 
behaviour with the same
resistance as in the normal state, $R_n$. ii) For smaller voltages, $ |V|<2\Delta/e$, the resistance is approximately $ R_n/2$ 
and exhibits subgap 
features.  iii) At the zero voltage $V=0$, the device switches to zero-resistance, exhibiting a Josephson current. 

In the next sections we perform quantitative analysis of the IVC, based on a detailed characterization of the normal state  current transport  in  the wire.

\section{Normal state transport}

In order to characterize the normal-state properties of the junctions, dc-measurements have been performed on several devices with a broad range of 
lengths and resistances. Measurement results for representative devices are sumarized in Table\,I. The devices are divided into three groups: A, B, and C
corresponding to the two-terminal, multi-terminal and suspended devices as shown in Fig.\,\ref{fig:SEM}. 
\begin{figure}[t]
\center
\includegraphics[width=0.9\linewidth]{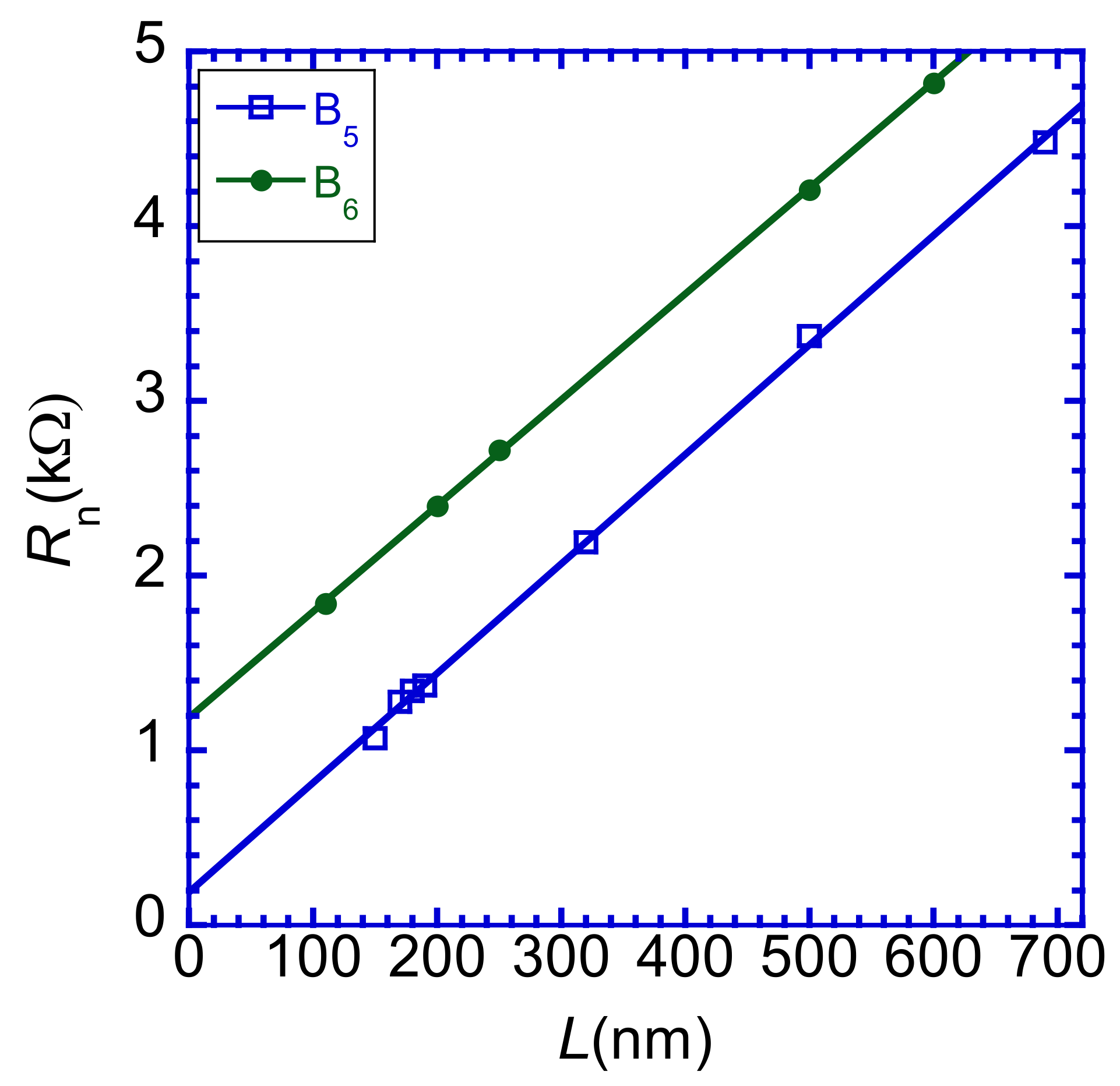}
\caption{Normal state differential resistance as a function of length for devices B$_5$ (blue line) and B$_6$ (green line). The two devices show approximately the same resistance per unit length 6\,$\Omega$/nm. Extrapolating to zero length gives the full contact resistance, less than 180\,$\Omega$ for B$_5$, and 1.2\,$k\Omega$ for  B$_6$.}
\label{fig:normalstateR}
\end{figure} 
%

The normal state resistance as a function of length for  devices B$_5$  and B$_6$ is plotted in Fig.\,\ref{fig:normalstateR}. 
The resistance of each device
increases linearly with length with approximately the same resistance per unit length, $R/L\approx6$\,$\Omega$/nm. Here, the resistance
values are taken  from the two-point measurements that also include  interface resistance. From the length dependence of the resistance in Fig
\,\ref{fig:normalstateR}, we extract the contact resistance by extrapolating to zero length. For device B$_5$,  we find that  contact  contribution is less than 180\,$\Omega$, while for device 
B$_6$, it is approximately 
$1.2$\,k$\Omega$. 

Taking advantage of multiple contacts of the B type devices we perform two- and four-point measurements of the resistance, which allows us to determine 
the number of conducting channels and the channel average  transparency.  The two- and four-point resistance expressions can be written for perfect interfaces as $R_{2p}
=R_q/(NT_t)$ and $R_{4p}=R_q(1-T_t)/(NT_t)$, respectively,\cite{datta} where $R_q=h/2e^2=12.9\,{\rm k}\Omega$ is the quantum of resistance, 
$N$ is the
number of channels and $T_t$ is the average transparency of the channels. Taking two- and four-point resistance measurements on the same section, we find that our nanowires have a spread in the number of channels ranging between 50 and 100 channels. For instance, for junction B$_{5c}$, we measured
 $R_{2p}=1.34\,k\Omega$, and $R_{4p}=1.18\,k\Omega$, giving the number of channels 
 \begin{equation}
N=\frac{R_q}{R_{2p}-R_{4p}}\approx80.
\end{equation}
This is consistent with the contact resistance found for device B$_5$, and implies perfect S-NW interfaces with transparency close to unity. 
For the same junction we can then extract the average transparency $T_t=0.12$ for the channels.
 
Assuming only a surface layer of nanowire to be conducting, in analogy with the 2DEG conductivity in planar InAs devices, such a large amount of conducting channels would 
give unrealistically small value for the Fermi wavelength. On the other hand, assuming the whole bulk of the wire to be conducting we find the electronic
Fermi wave length $\lambda_F$ to depend logarithmically on the number of channels,  as  
\begin{equation} 
\lambda_F(N)=\frac{2 \pi r_w}{\alpha_{l,n}}\approx A-B\,{\rm ln}(N).
\label{numberofmodes}
\end{equation}
This result is arrived at by solving the Schr\"odinger equation in a cylinder of radius $r_w$ and 
counting the number of modes (channels) that cross the Fermi level. In Eq.~(\ref{numberofmodes}) $\,\,
\alpha_{l,n}$ is the $n$-th zero of the $J_l$-th Bessel function, $l$ labels the last mode
that contributes to transport. The coefficients  $A$ and $B$ in Eq.~(\ref{numberofmodes}) are functions of the nanowire radius; for  $r_w=40\,{\rm nm}$,
 $A=42.4\, {\rm nm}$ and $B=7.63\, {\rm nm}$,  and varying the radius by 
$\pm10\%$ will change both coefficients approximately by $ \pm 2\%$. 
We can bracket the Fermi wave length between, $\lambda_F \approx 17\,{\rm nm}$ (100 channels) and $\lambda_F \approx22\, {\rm nm}$ (50\, channels) for
$r_w=40\, {\rm nm}$.
Our values are consistent with the ones reported for planar InAs 2DEG ($\lambda_F \approx 18\, {\rm nm}$), \cite{Peter} 
and for InAs nanowires ($22\, {\rm nm}\lesssim \lambda_F \lesssim 33\, {\rm nm}$).\cite{Jespersen}
In the further discussions we adopt the values, $\lambda_F=22\, {\rm nm}$, and $N=$55 for all the junctions. 

\begin{table}
    \begin{tabular}{ l c | c }
        Fermi wavelength $\lambda_F$&${\rm (nm)}$ & $22$\\
        \hline
        Fermi wavevector $k_F$& ${\rm (m^{-1})}$ & $2.9\cdot 10^8$\\
        \hline
        Fermi velocity $v_F$&${\rm (m/s)}$ & $1.3\cdot10^6$\\
        \hline
        Mean free path $\ell_{\rm e}$&${\rm (nm)}$ & $46$\\
         \hline
        \# conducting channels && $55$\\
        \hline 
         Superconducting gap $\Delta$&$ (\mu {\rm eV})$ & $160$\\
        \hline 
        Clean coherence length $\xi_0$&${\rm (nm)}$ & $1300$\\
        \hline
        Diffusive coherence length $\xi_D$&${\rm (nm)}$ & $245$\\
        \hline
        Diffusion constant $D_{\rm diff}$&${\rm (cm^2/s)}$ & $200$\\
        \hline
        $e\Delta/\pi \hbar$&${\rm (nA)}$ & 12.6

    \end{tabular}
    \caption{Summary of the extracted parameters for the nanowires.}
    \label{parameters}
\end{table}

Furthermore we use the measured resistance per unit length, $R/L=6$\,$\Omega$/nm, together with the expression for the Drude conductivity,  $\sigma=ne^2\tau/m$,  
to evaluate a mean free path for the nanowires of $\ell_{\rm e}=46$\,nm. The corresponding Fermi velocity 
$v_F=\hbar k_F/{m^*}\approx1.3\times10^6$\,m/s is evaluated by 
using an electronic
effective mass  ${m^*}=0.026m_e$ of bulk InAs, where $m_e$ is the free electron mass. The effective mass of electrons ${m^*}$ for planar InAs 2DEG has 
been estimated to be in a range from 0.024 to $ 0.04\times m_e$.\cite{Peter, Furusaki} The normal state properties of the nanowires are summarized in Table\,II.
%

\section{Theoretical model}

\begin{figure}[ht]
\center
\includegraphics[width=0.75\linewidth]{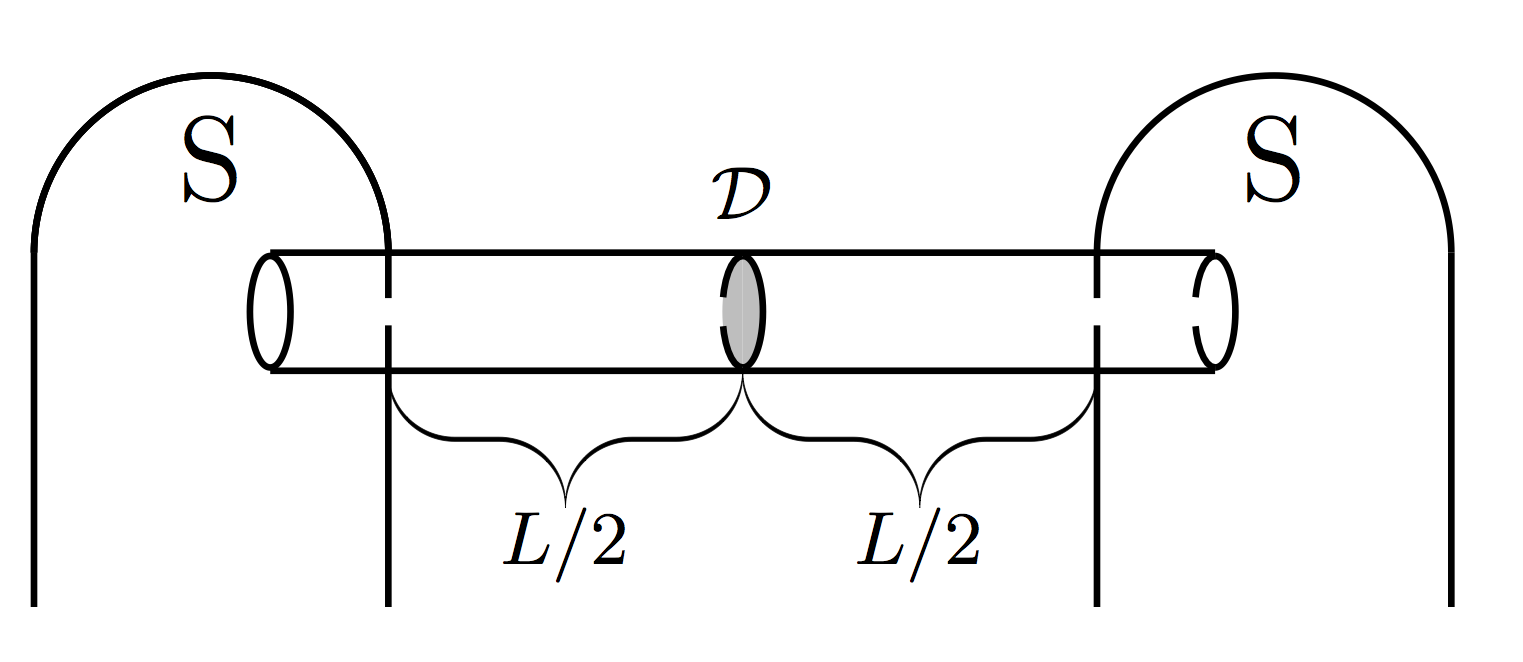}
\label{fig:device_sketch}
\caption{A schematic picture of the theoretical model for the nanowire junctions: Superconducting electrodes (S) are connected by a disordered nanowire of length $L$, which also contains crystalline defects; the latter are modeled with a lumped scatterer situating in the middle of the wire and having effective transparency $\mathcal{D}$; the applied voltage is assumed to mostly drop at the scatterer. }
 \label{nanowire}
\end{figure}

The major difficulty for theoretical interpretation of the experimental data is the very large spread of the wire lengths. 
Indeed, the shortest wire (30 nm) is in the ballistic point contact regime ($L<\ell_{\rm e},\xi_0$, with $\xi_0=\hbar v_f/2\pi k_BT_c$), while the longest wire is in the 
diffusive long junction regime ($L>\ell_{\rm e},\xi_D$, with $\xi_D=\sqrt{\ell_{\rm e}\xi_0}$). 
The majority of the junctions are in the intermediate cross over region.  
Furthermore, all the tested junctions exhibit the Josephson effect. This implies that the current transport is fully coherent and requires theoretical modeling within the framework of the coherent MAR theory.\cite{Arnold1987,Bratus1995,Averin1995,Cuevas1996}
To overcome this difficulty, we adopt a
simple and tractable model, with which we can bridge between the ballistic and diffusive transport regimes, and describe the cross over to the long junction behavior.\cite{comment} 
The model setup is shown in Fig.~\ref{nanowire}. We assume that the two superconducting leads are connected to the nanowire by highly transmissive contacts, which are
treated as fully transparent. The nanowire is disordered due to elastic scattering by impurities and crystal imperfections.
This is treated in the Born approximation and the mean free path estimated from the experiments $\ell_{e}=46\,{\rm nm}$ infers a scattering rate 
$\Gamma=v_F/\ell_{\rm e}\approx 2.8\cdot 10^{13}{\rm s}^{-1}$. Strong defects are included and treated as a single interface having the same effective transparency
${\mathcal{D}}$ for all conducting channel. This defect is assumed to be in the centre of the nanowire, and the applied voltage to drop at the defect. 

Using the quasi-classical Green's function methods described in Refs.~\onlinecite{Cuevas2001,Eschrig2009}, 
we calculate the IVC as function of device length and transparency by solving the coherent 
MAR problem.  The current is calculated at the scatterer\cite{Cuevas2001} and expressed through the boundary values, $\hat g_{R/L} = \hat g(\hat {\vec p}_F,x=\pm0;\omega)$, of the quasi-classical Green's function for a given channel,
\begin{equation}
\hat g(\hat {\vec p}_F,x;\omega)=
\left(\begin{array}{lr}g(\hat {\vec p}_F,x;\omega) &f(\hat {\vec p}_F,x;\omega)\\ \tilde f(\hat {\vec p}_F,x;\omega)&-g(\hat {\vec p}_F,x;\omega) \end{array}\right),\,\,\hat g^2=-\pi^2.
\label{gf}
\end{equation} 
The Green's function is computed by solving the Eilenberger equation in the right and left parts of the nanowire, 
\begin{equation}
i \hbar\vec v_F\cdot \vec \partial_x \hat g(\hat {\vec p}_F,x;\omega)+[\varepsilon(x;\omega)\hat \tau_3-\hat \Delta_{\rm{imp}}(x;\omega),\hat g(\hat {\vec p}_F,x;\omega)]=0,
\label{eilenberger}
\end{equation} 
complemented with the Zaitsev boundary conditions at the scatterer and NW-S interfaces.\cite{Zaitsev1984,Cuevas2001,Eschrig2009} $\hat \tau_3$ is the third Pauli matrix in Nambu space. In Eq.~(\ref{eilenberger}) we introduce the impurity scattering via the impurity self-energies
\begin{eqnarray}
\varepsilon(x;\omega)&=&\hbar\omega-\hbar\Gamma \langle g(\hat {\vec p}_F,x;\omega)\rangle_{p_F},\\
\hat \Delta_{\rm{imp}}(x;\omega)&=&\hbar\Gamma \langle \hat f(\hat {\vec p}_F,x;\omega)\rangle_{p_F},
\end{eqnarray}
$\langle \cdots\rangle_{p_F}$ is average over directions ($\pm \hat {\vec p}_F$).
The matrix $\hat f$ is the anomalous (off-diagonal) part of the Green's function $\hat g$. The components $(f,\tilde f)$ of $\hat f$ describe
the pairing correlations leaking in to the nanowire and two are related by symmetry as $\tilde f(\hat {\vec p}_F,x;\omega)=-f^*(-\hat {\vec p}_F,x;-\omega^*)$.

\section{Excess current } 

\begin{figure}[ht]
\center
\includegraphics[width=0.75\linewidth]{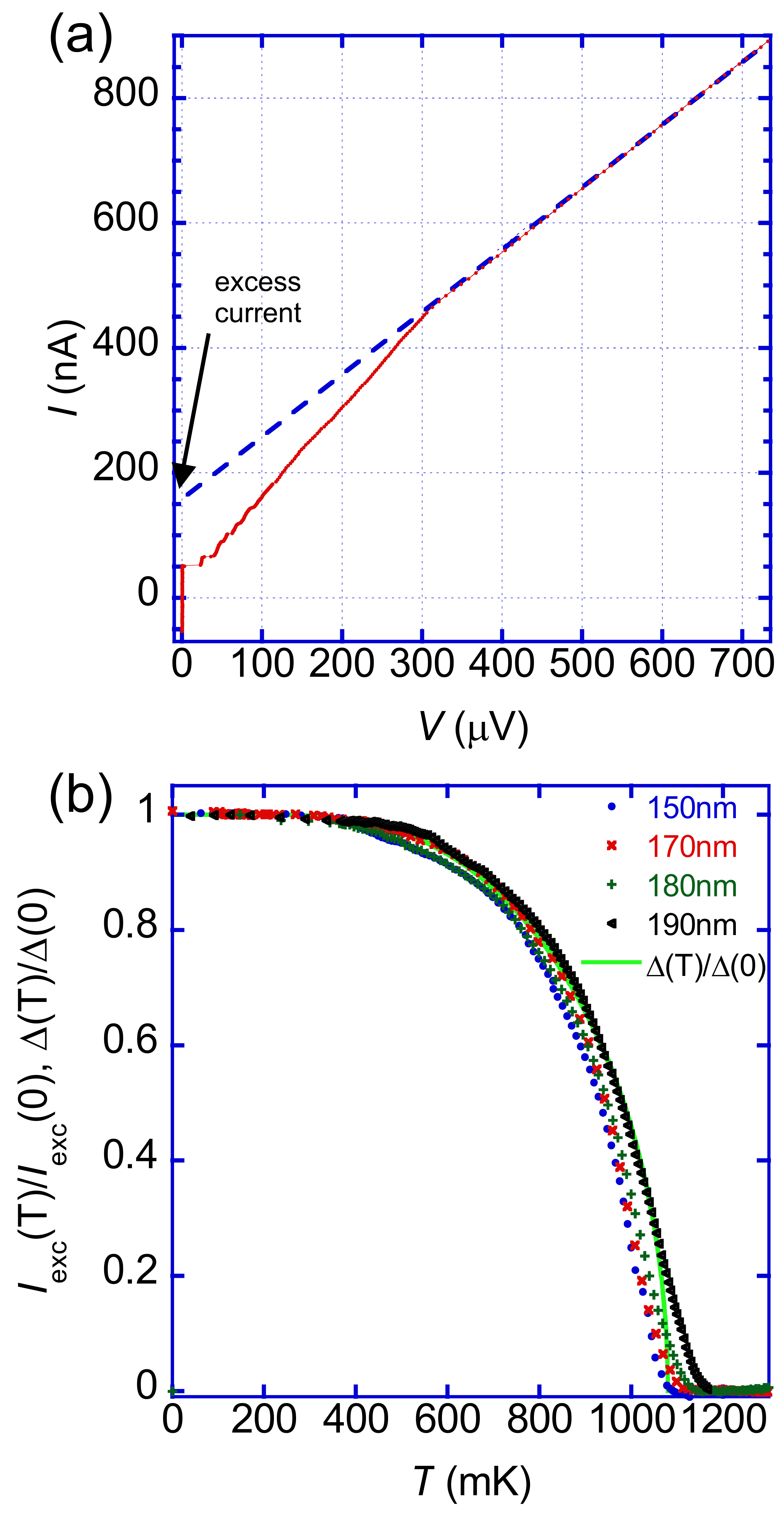}
\caption{a) Current-voltage characteristics of the device B$_{5a}$ with length $L=150$\,nm and normal state resistance $ R _{n}=1.07$\,k$\Omega$. 
The excess current is extracted by extrapolating the IVC from high voltage to zero voltage. 
b) Excess currents as a function of temperature are shown for device B$_{5}$ (L$=150$\,nm, 170\,nm, 180\,nm, and 190\,nm). The excess 
currents follow the superconducting energy gap $\Delta(T)$ (light green).}  
\label{fig:excess}
\end{figure}

We start with a discussion of the excess current at large voltage, which is a robust feature of the proximity IVC. The excess current, $I_{\rm exc}$, is extracted from the current-voltage characteristics at large voltage bias using
the asymptotic form 
$I(V> 2\Delta/e)\approx V/R_N+I_{\rm exc}+\mathcal{O}(\Delta/eV)$. The excess current contains contributions  both from the  single-particle and from the two-particle Andreev currents, and it linearly scales with the energy gap $\Delta(T)$ (see {\em e.g.} Ref.~\onlinecite{Shumeiko1997}).
In Fig.\,\ref{fig:excess}a, the excess current is obtained for the device $B_{5a}$ by extrapolating a linear fit of the IVC measured at $V> 2\Delta/e$ (blue dashed line) giving $I_{exc}=150$\,nA.
 To  verify that the measured excess current derives from Andreev scattering processes the experimentally extracted excess
current is plotted as a function of temperature in Fig. \ref{fig:excess}b. As can be seen the amplitude of the excess current follows the temperature dependence of the superconducting gap $\Delta(T)$. 

\begin{figure}[ht]
\center
\includegraphics[width=0.99\linewidth]{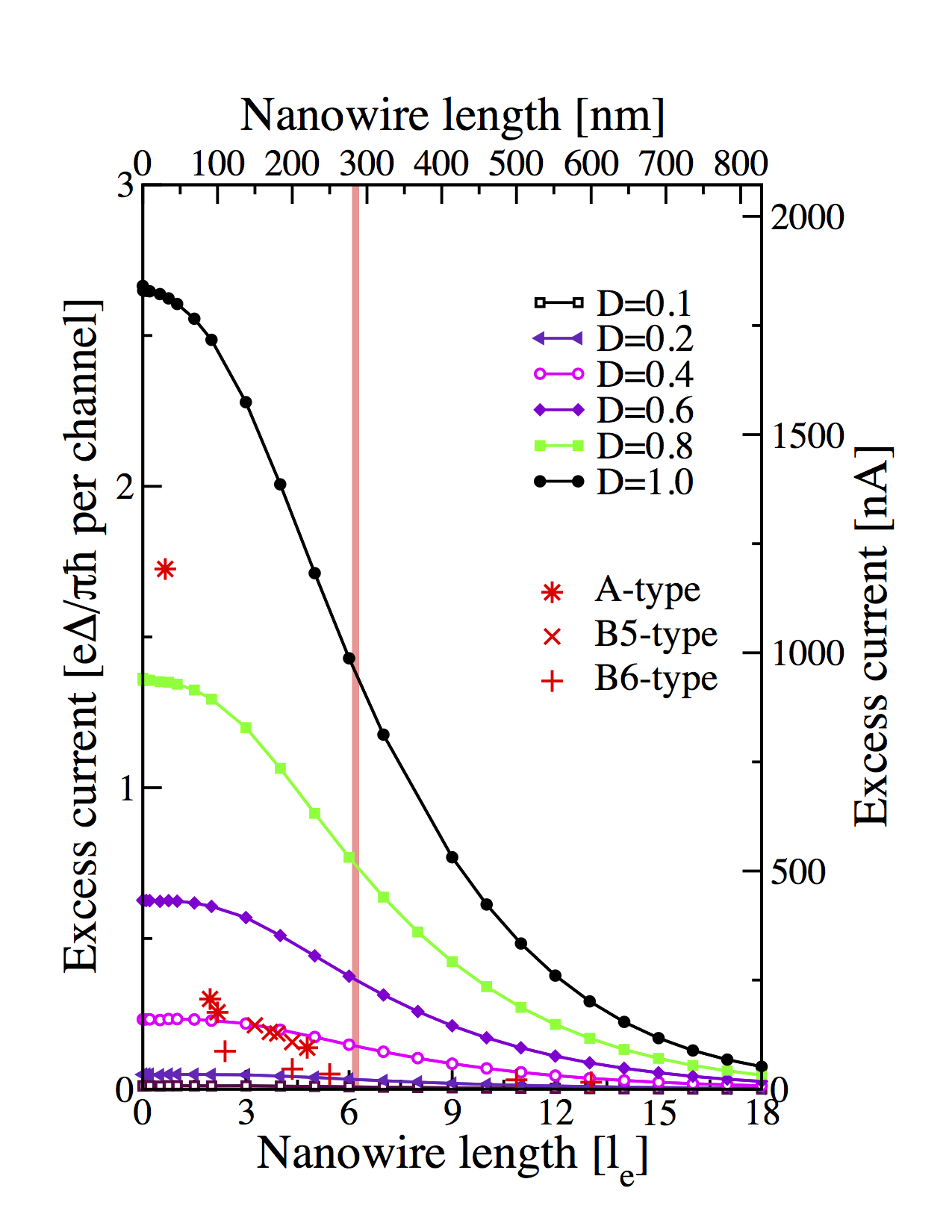}
\caption{The computed excess current as a function of length and effective transparency of the device. The units of the excess current are given on the left axis 
for the single channel and to the right in nA assuming that all channel have the same average transparency. The length of the device is given both in units of the 
mean free path (bottom x-axis) and in nm (top x-axis). The vertical line indicates the length $L=\sqrt{\hbar D_{\rm diff}/\Delta}\approx 1.1 \xi_D$, 
where the Thouless energy, $E_{\rm Th}=\hbar D_{\rm diff}/L^2$, equals the superconducting gap; 
this length separates the short-junction limit ($E_{\rm Th}\gg\Delta$) from the long-junction limit ($E_{\rm Th}\ll\Delta$).  
The stars and crosses are the experimental data from Table I; as can be
seen the most devices are in the intermediate limit where $E_{\rm Th}\approx\Delta$.
} 
\label{fig:Iexc_vs_D_and_L}
\end{figure}

The excess current also depends on the transparency and the length of the nanowire device.  
In Fig.~\ref{fig:Iexc_vs_D_and_L} we present the computed excess current as function of device length together with $I_{\rm exc}$ extracted from the measurements.
The maximum values of the theoretical curves correspond to the point contact limit ($L=0$), and they are in a good agreement with analytical results\cite{Shumeiko1997}, 
$I_{\rm exc}=({8}/{3\pi})( {e\Delta}/{\hbar})$ per channel for $\mathcal{D}=1$, and $I_{\rm exc}\approx \mathcal{D}^2 ({e\Delta}/{\hbar})$ for $\mathcal{D} \ll 1$. When the wire length is increased the excess current decreases. This was also found, experimentally and theoretically, in ballistic 2DEG InAs,\cite{Peter}, and computed for fully diffusive junctions.\cite{Cuevas,Bezuglyi} In our case, the experimental values fall on curves with a typical effective transparency between $0.2$ and  $0.4$ being only weakly device dependent between batches of nano wires. 
These values compare favorably with $T_t=0.12$ extracted from the 2-point and 4-point measurements in the normal state. 

One device, $A_1$ (L=30\,nm), however, stands out 
showing a high transparency of $\mathcal{D}\approx 0.87$. For this junction, highly transmissive ballistic point contact, one should anticipate the largest critical current.

\section{Josephson current}

Next, we discuss the Josephson critical current  as a function of length, temperature, and magnetic field.
The maximum values of the Josephson current, $I_{m }$, are extracted from the experimentally obtained IVC at the base temperature of 15\,mK and shown in Table I.  
The maximum currents exhibit a range of values depending on the resistance and length of the devices, from a few nA to 800\,nA. 
Similarly, the characteristic voltage, the $I_{m}R_{n}$-product, also exhibits a range of values, from $20\,\mu$V to $130\,\mu$V.

\begin{figure}[ht]
\center
\includegraphics[width=0.99\linewidth]{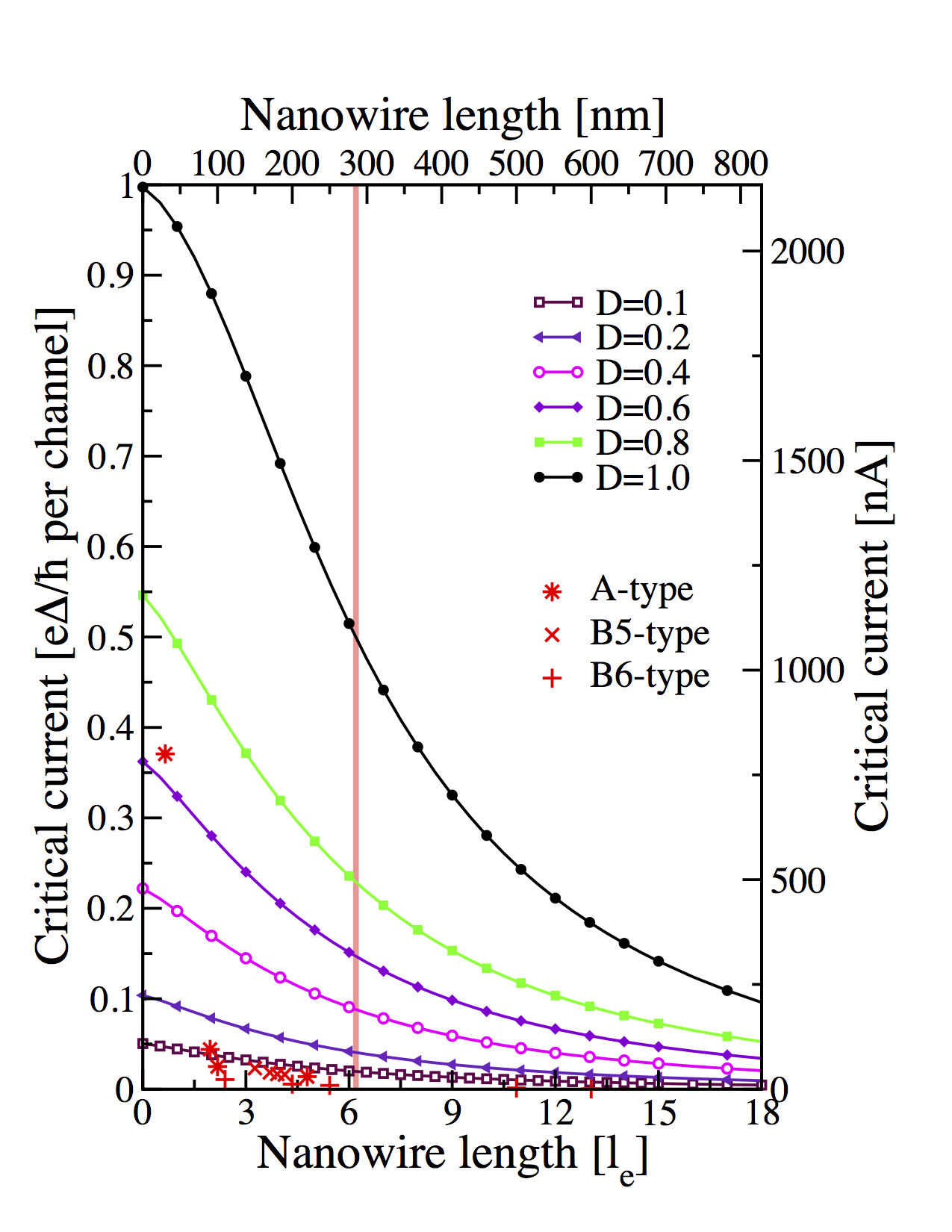}
\caption{The computed critical current $I_c$ as a function of length and effective transparency of the device. The units of the maximum current are given on the left axis 
for the single channel and to the right in nA assuming that all channel have the same average transparency. The length of the device is given both in units of the 
mean free path (bottom x-axis) and in nm (top x-axis). The single markers are the experimental values of the maximum Josephson current $I_m$ reported in Table I.
}  
\label{fig:Ic_vs_D_and_L}
\end{figure} 

Theoretically, the Josephson current-phase relation  is computed using boundary values of the Green's function, $\hat g(\hat {\vec p}_F,x;\omega)$, in Eqs.~(\ref{gf}) and (\ref{eilenberger}), 
the expression reads,\cite{Zaitsev1984}
%
\begin{gather}
    I_s(\phi)=\frac{8\pi eT\mathcal{D}}{h}\sum_{\omega_n>0}\Bigg\langle\frac{f_R  f_L \sin\phi}{2\!-\mathcal{D}(g_R g_L \!-\!f_R f_L \cos\phi+\!1)}\Bigg\rangle_{p_F}.
    \label{Is}
    \end{gather}
The sum is over all Matsubara frequencies, $\omega_n=\pi k_BT(2n+1)$,  $T$ is the temperature,  $\phi$ the phase difference over the junction. 
The critical current is  obtained by maximizing the supercurrent.

The maximum Josephson currents presented in Table I, together with a theoretical critical current fit, as a function of length, are plotted in Fig.\,\ref{fig:Ic_vs_D_and_L}. The shortest junction exhibits the largest Josephson current, as expected, with the theoretical fit of the transparency, $\mathcal{D}\approx 0.65$. This is very close (75\%) to the theoretical limit defined by the transparency extracted from the analysis of the excess current. 
The other junctions fall in the transparency region $0.05 < \mathcal{D} < 0.1$, which is smaller (approximately by a factor of 4) compared to the transparency 
extracted from the excess current. 

Similar or even larger reduction of Josephson current is commonly observed in nanowires, and it is also
common in 2DEG InAs Josephson junctions\cite{Peter}. Such an effect  is  not well understood, perhaps it could be related to some depairing mechanism, for example due 
to magnetic scattering. 

One would expect a certain suppression of the Josephson current extracted from the IVC measurement compared to the equilibrium critical current 
due to the effect of phase fluctuations. However, our analysis shows that majority of our junctions are 
overdamped, and the suppression 
of the critical current in this regime is relatively small and cannot account for the whole suppression effect. Indeed, the capacitances of the devices 
are estimated in the range,  $C\sim$1-5\,fF, cf. Ref.~\onlinecite{Abay}. Assuming  $C=5$ fF and the junction 
resistance, $R_0 \sim 100 \Omega$ at plasma frequency $\sim$ 1 GHz corresponding to the free space impedance, we estimate the quality factor $Q=\sqrt{2eI_c C/\hbar}R_0 \lesssim 0.1$ for the representative junction with critical current, $I_m = 50$ nA. This estimate refers to an unbiased junction, the Q-factor further decreases when the current bias is applied. 
For such an overdamped regime, $Q\ll1$, the switching probability is significantly suppressed \cite{Martinis1988}, and IVC can be modeled with the 
Ambegaokar-Halperin theory.\cite{Ambegaokar1969} This conclusion is supported by the absence of hysteresis on IVC. The  IVC measurement  takes approximately
1 minute, so that the sample spends approximately few seconds close to Ic.
Assuming the temperature of electromagnetic fluctuations being close to the base temperature of 15 mK due to a careful noise filtering,\cite{Bladh2003} we find that the suppression 
effect accounts for approximately 20\% of the theoretical value for the majority of the junctions with critical currents exceeding 10 nA. For the shortest junction with $I_m=800$ nA  the suppression is even smaller, about few percent.

\begin{figure}[t]
\includegraphics[width=0.99\linewidth]{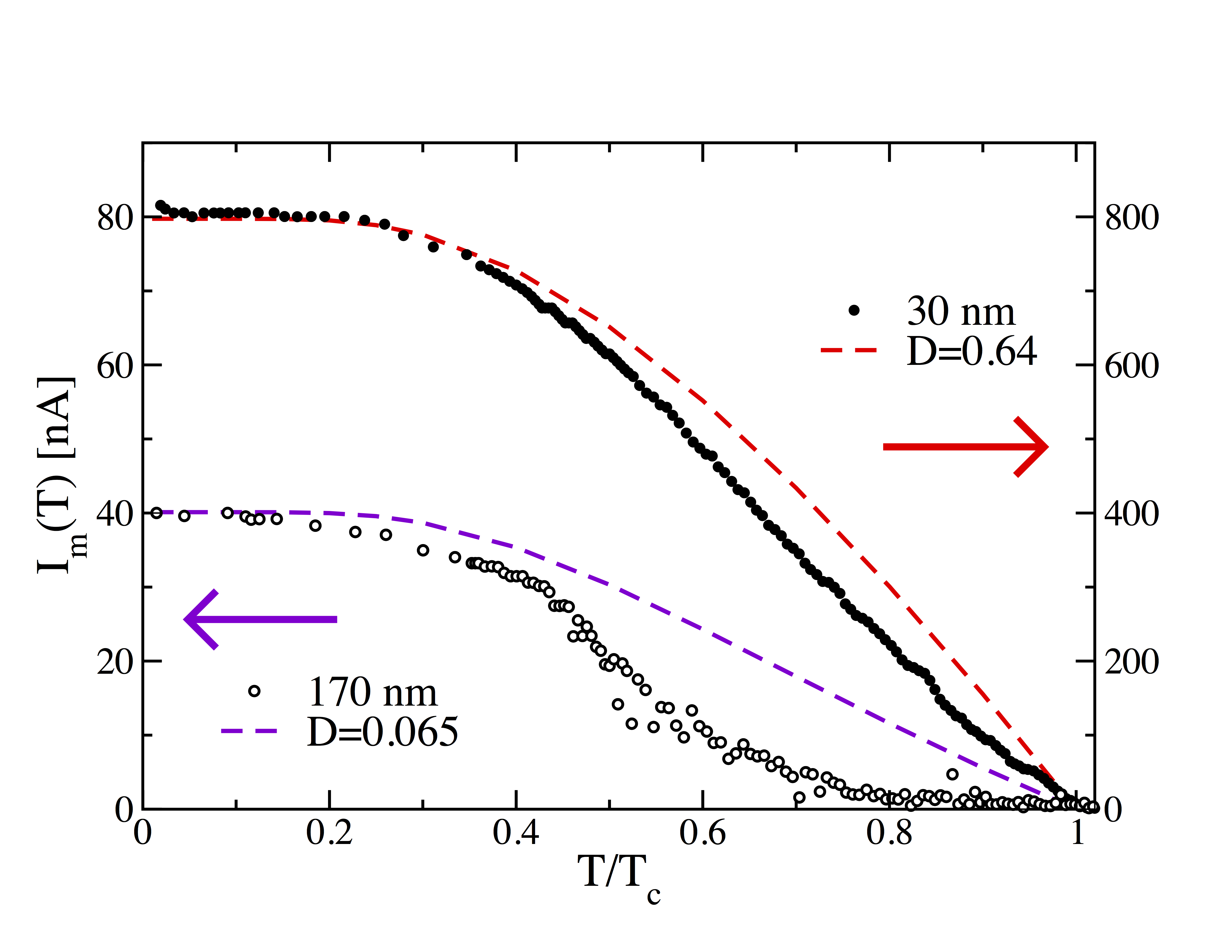}
\caption{
Maximum Josephson current as a function of temperature for two devices of length $L=30$\,nm and $L=170$\,nm. Note the different scales for the current magnitude. 
Theoretical fits to the critical current for both devices are shown. The transparency chosen to fit the data is taken from the low-temperature values for $I_m$ in Fig.~\ref{fig:Ic_vs_D_and_L}. 
}
\label{fig:Ic(T)}
\end{figure} 

The maximum Josephson current is also investigated as a function of temperature for several devices. The maximum currents for the shortest length device 
A$_1$ ($L\approx30\,$nm and $R_{n}= 0.16$\,k$\Omega$), and for the somewhat longer device B$_{5b}$, ($L\approx170\,$nm and $R_{n}= 1.15$\,k$\Omega$), 
are shown in Fig.\,\ref{fig:Ic(T)}. At the base temperature $T=$15\,mK, the devices have maximum Josephson currents of $I_{m}=800$\,nA and 40\,nA, respectively.
The data for the shortest device agree well with theory in a broad range of temperatures. The longer device exhibits a concave shaped decay at higher 
temperatures and deviates from the theoretical fit. Qualitatively similar shape of $I_c(T)$ has been theoretically found for diffusive junctions with highly resistive 
interfaces (SINIS),\cite{Kupriyanov1999} and explained with enhancement of electron-hole dephasing in the proximity region due to large dwell time. Such an effect 
is similar to the effect of increasing length of the junction ({\em cf.} Ref.~\onlinecite{Bezuglyi}). Given such a similarity we may conclude that although device B$_5$ has
transparent S-NW interfaces, the model\cite{Kupriyanov1999} might better capture the effect of the junction length. 
\begin{figure}[t]
\includegraphics[width=1.0\linewidth]{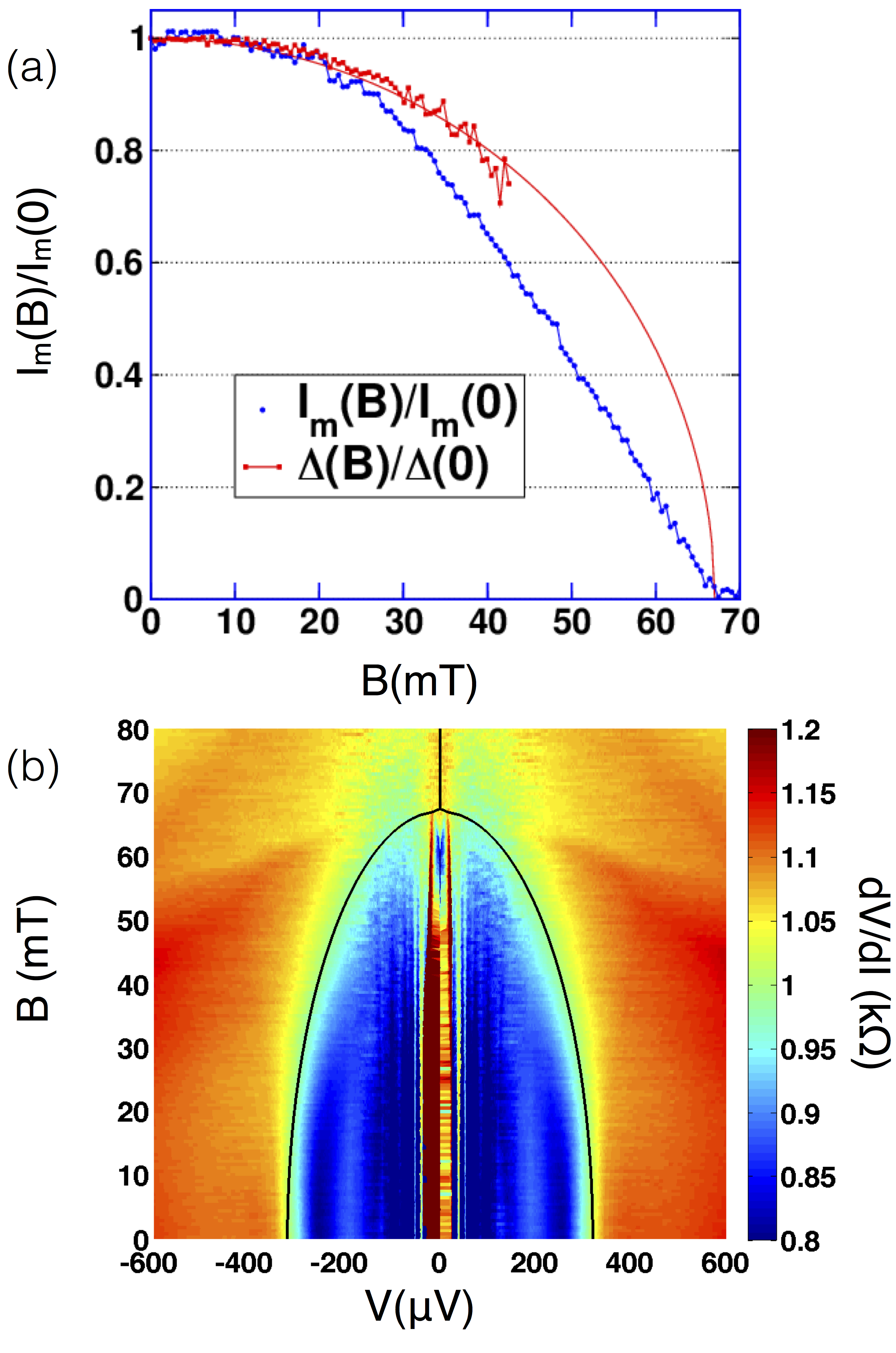}
\caption{Panel (a): Normalized maximum Josephson current (blue dots) as a function of magnetic field for device B$_{5b}$ ($L=170$\,nm) 
along with the extracted superconducting energy gap (red dots). The redline is $\Delta(B)=\Delta(0) \sqrt{1-(B/B_c)^2}$. This fit (black line) is made to the 
gap-like feature in the $dV/dI (V,B)$-data displayed in panel b.}
\label{fig:Ic(B)}
\end{figure} 

At the base temperature of 15\,mK, we also have obtained IVCs as a function of magnetic field. The magnetic field is applied perpendicular to the superconducting 
leads. The normalized maximum Josephson current and the superconducting energy gap as a function magnetic field are plotted in Fig.\,\ref{fig:Ic(B)} for device B$_{5b}$ with  
$L=170$\,nm. The superconducting gap $\Delta(B)$ is fitted to the expression $\Delta(B)=\Delta(0) \sqrt{1-(B/B_c)^2}$ from which we extract $B_c=67$\,mT. 
The maximum current decreases and is 
totally suppressed above $B_c$. No Fraunhofer oscillations are observed in any of the devices, consistent with a suppression of 
superconducting energy gap in the leads. 

\section{Subgap current}

\begin{figure}[t]
\includegraphics[width=0.8\linewidth]{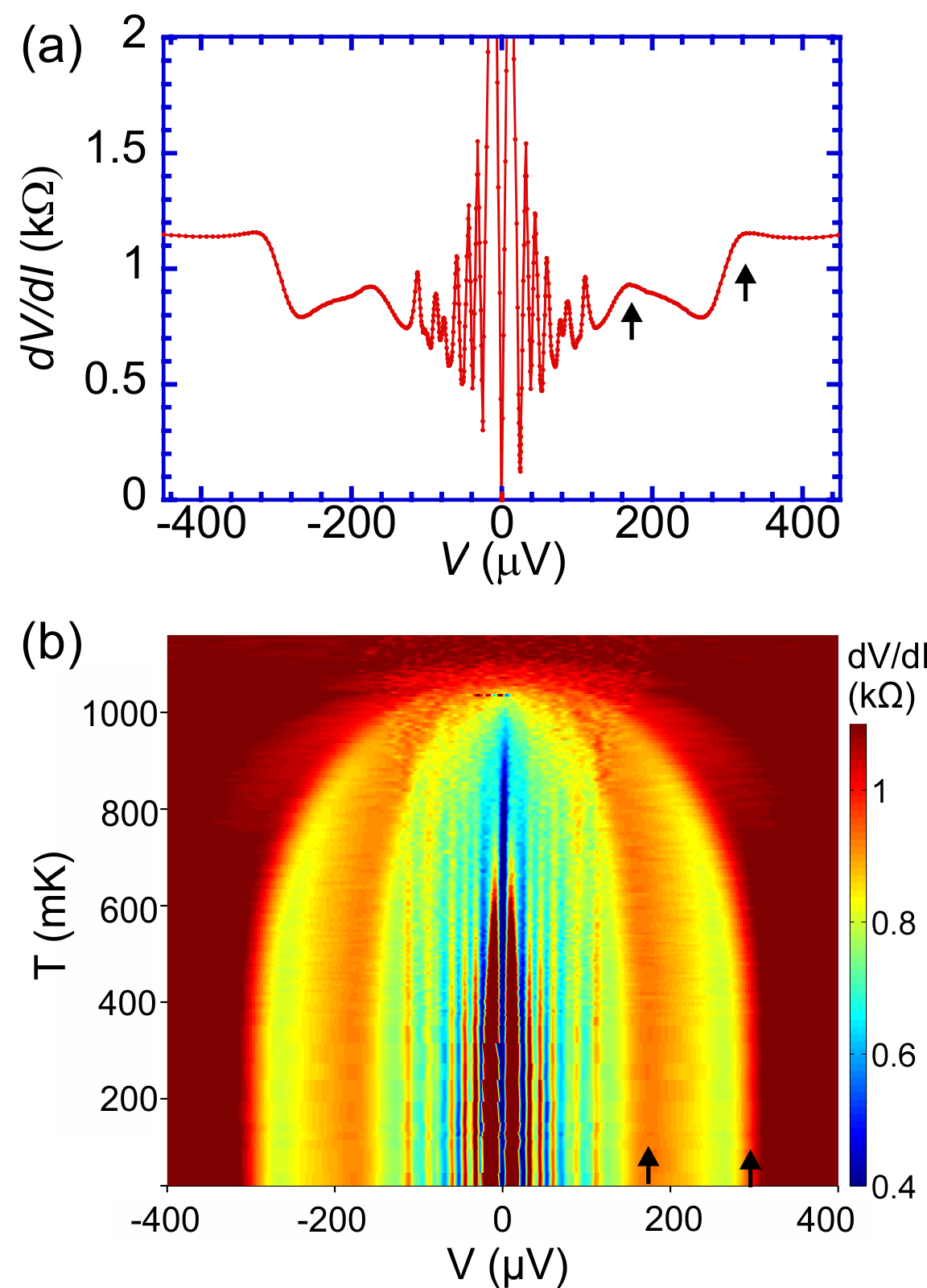}
\caption{ (a) Differential resistance $dV/dI$ as function of voltage for device  B$_{5b}$ ($L=170$\,nm and  $R_{n}=$1.15\,k$\Omega$). 
The resistance substantially decreased at $V\approx2\Delta/e$ and exhibits symmetric resistance peaks/dips. (b) An image plot of differential resistance $dV/dI$ 
as a function of  voltage and temperature. As the temperature is increased, the first two peaks, marked by two arrows, smoothly move towards lower voltages
consistent with the decrease of the superconducting energy gap $\Delta(T)$. However, the voltage positions of the other peaks are independent of temperature.}
\label{fig:subgap_vs_T}
\end{figure}

Now we proceed with discussion of the IVC in the subgap region, $V < 2\Delta/e$, as function of temperature and magnetic field, and for different nanowire 
lengths. A typical plot of the differential resistance as a function of voltage is presented in Fig.\,\ref{fig:subgap_vs_T}a. The resistance drops from 
$R_{n}=1.15$\,k$\Omega$ at $V\gg 2\Delta$ to $R_{SG}\approx0.7$\,k$\Omega$ at $V\approx 260\,\mu$V, which corresponds to the gap value, 
$2\Delta/e$. Such a drop of resistance in the subgap region is a characteristic of Andreev transport in transmissive SNS junctions.\cite{Blonder1982}
%
%
Furthermore, the differential resistance shows a second feature at approximately half the gap voltage, $V\approx130\,\mu$V$=2\Delta/2e$  
(shown by arrow in Fig.\,\ref{fig:subgap_vs_T}a).

Positions of both these features scale with the temperature dependence of the superconducting gap $\Delta(T)$, as shown on Fig.\,\ref{fig:subgap_vs_T}b.  
This unambiguously indicates the MAR transport mechanism. Similar features associated with MAR are observed in all measured 
devices, in some devices we also observed a third MAR feature at $2\Delta/3e$.

\begin{figure}[h]
\includegraphics[width=0.8\linewidth]{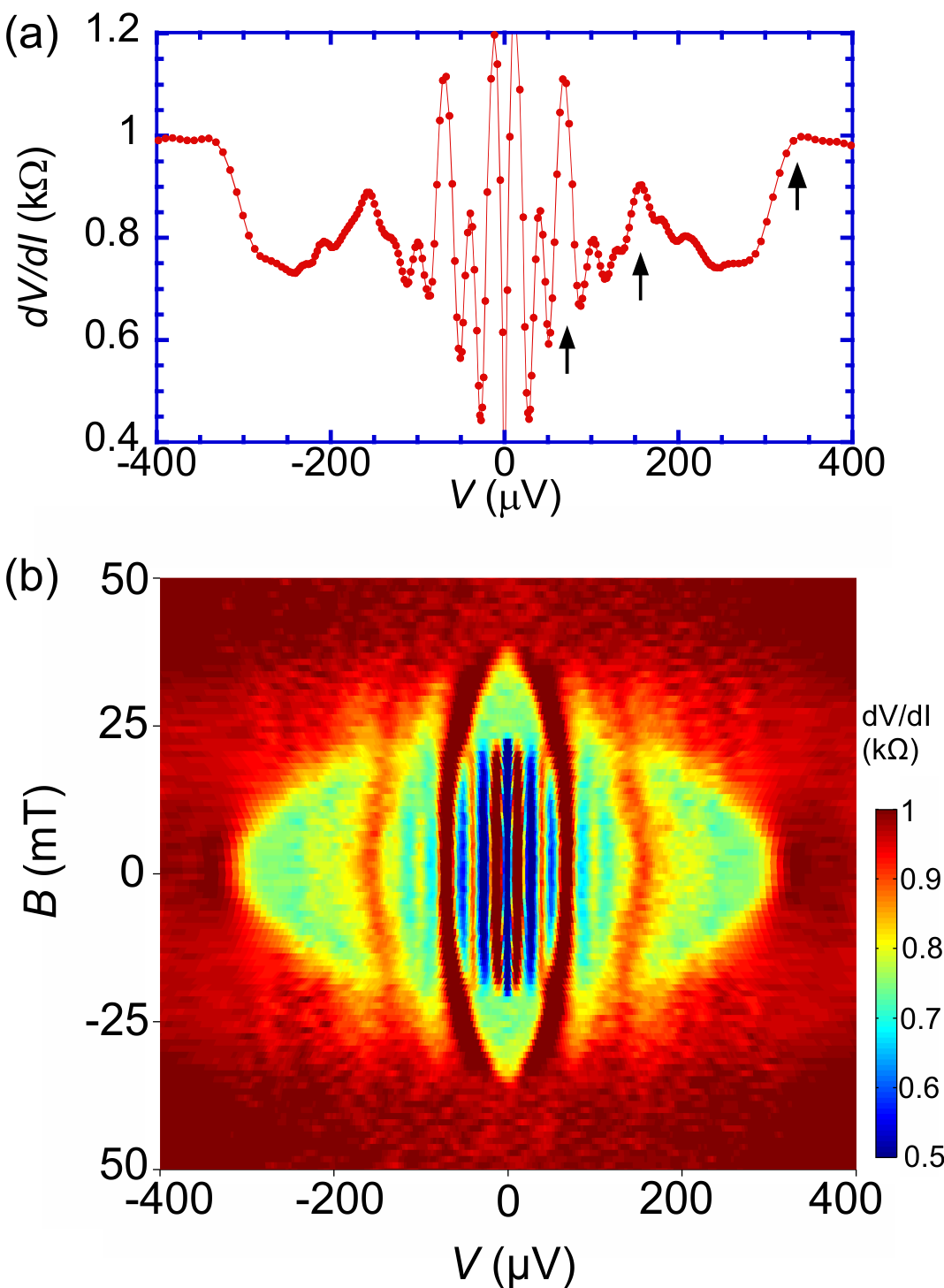}
\caption{(a) Differential resistance  $dV/dI $ as a function of voltage for device A$_4$. There are three MAR features indicated with arrows. (b) An image plot of differential resistance $dV/dI$ 
as a function of  voltage and magnetic field. As the magnetic field is increased, the MAR resistance peaks smoothly move towards lower voltages consistent with the decrease of the superconducting energy gap $\Delta(B)$. However, the voltage positions of the other peaks are independent of magnetic field.}
\label{fig:dV_dI_B}
\end{figure}

We have also measured the dependence of positions of the resistance features as a function of magnetic field. The differential resistance of device $A_4$ as function 
of magnetic field is shown in  Fig.\,\ref{fig:dV_dI_B}. In this device, the three MAR features are present (marked by arrows), which move smoothly 
towards lower voltages following the magnetic field dependence of the  gap $\Delta(B)$.

\begin{figure}[h]
\includegraphics[width=0.75\linewidth]{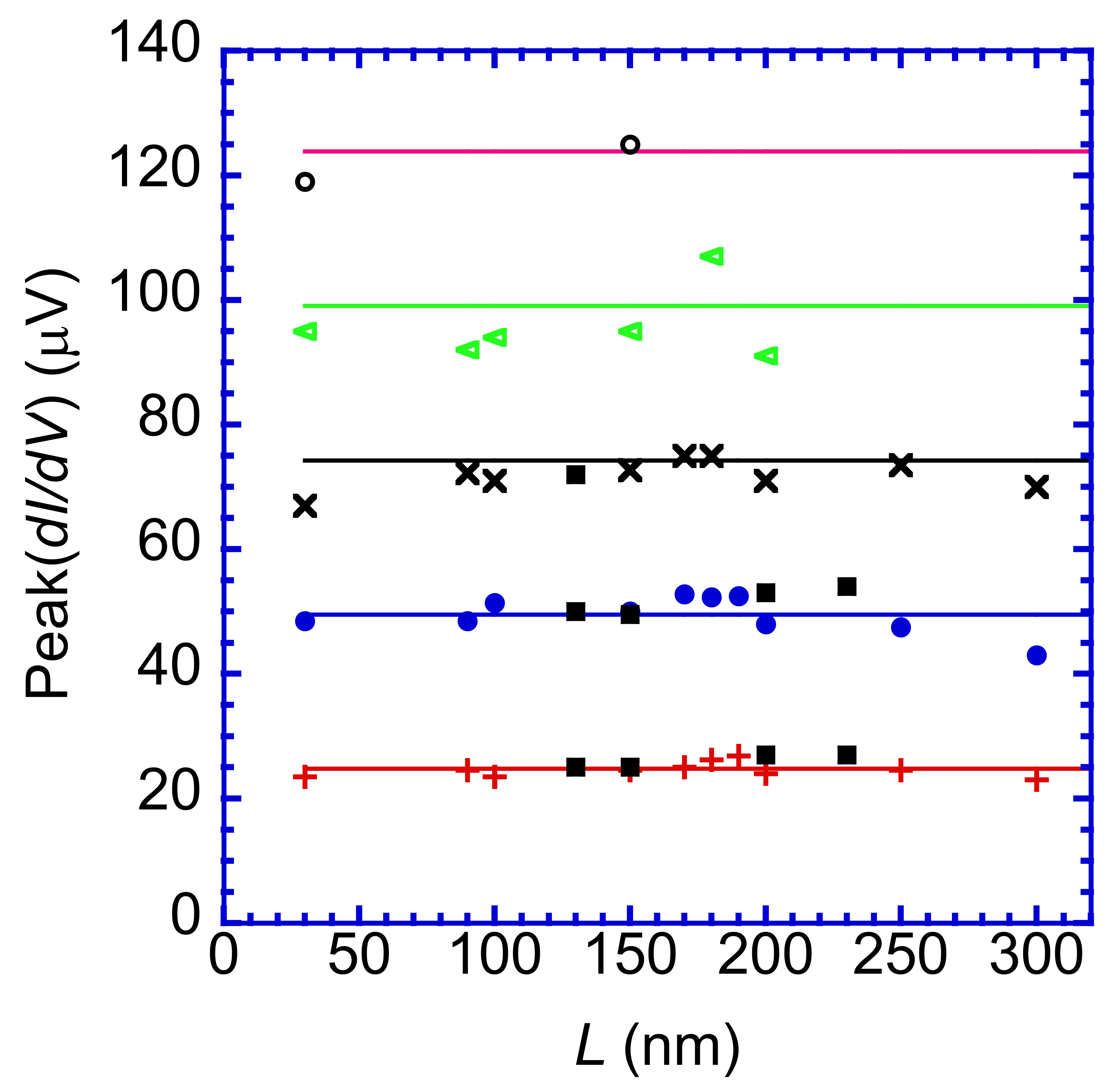}
\caption{ Voltage positions of the temperature independent differential conductance peaks as function of length for  devices of type A, B and C. Horizontal lines indicate the multiples of the voltage $V\approx 24\mu$V.
}
\label{fig:peaks}
\end{figure}

Besides the MAR features, the IVC of all the measured devices exhibit a number of structures at lower voltages, whose positions are independent of both
temperature and magnetic field, see Fig.\,\ref{fig:subgap_vs_T} and  Fig.\,\ref{fig:dV_dI_B}. These structures are therefore not associated with MAR. However, 
they are related to the superconducting state in the electrodes since they do not persist above the critical temperature and critical magnetic field and 
even  disappear somewhat earlier. 

The origin of these structures is not clear. In Ref.~\onlinecite{Andrey}  similar structures were reported for suspended NW devices and attributed to resonances 
resulting from
coupling of the ac Josephson current to mechanical vibrations in the wire. The fact that we observe such structures not only in suspended but also in non-suspended
wires rules out this explanation. Furthermore, the phonon resonances would appear at voltages corresponding to the phonon eigen-frequencies, i.e. depend on the wire length
($\propto 1/L$).  We systematically measured the length dependence of the low-voltage, temperature independent structures, the results are
presented in Fig. \ref{fig:peaks}. The positions of the structures do not depend on the wire lengths neither for suspended nor non-suspended devices. The positions are given by 
the integer multiples of the same voltage, $V\approx 24\,\mu$V.

\begin{figure}[h]
\includegraphics[width=0.75\linewidth]{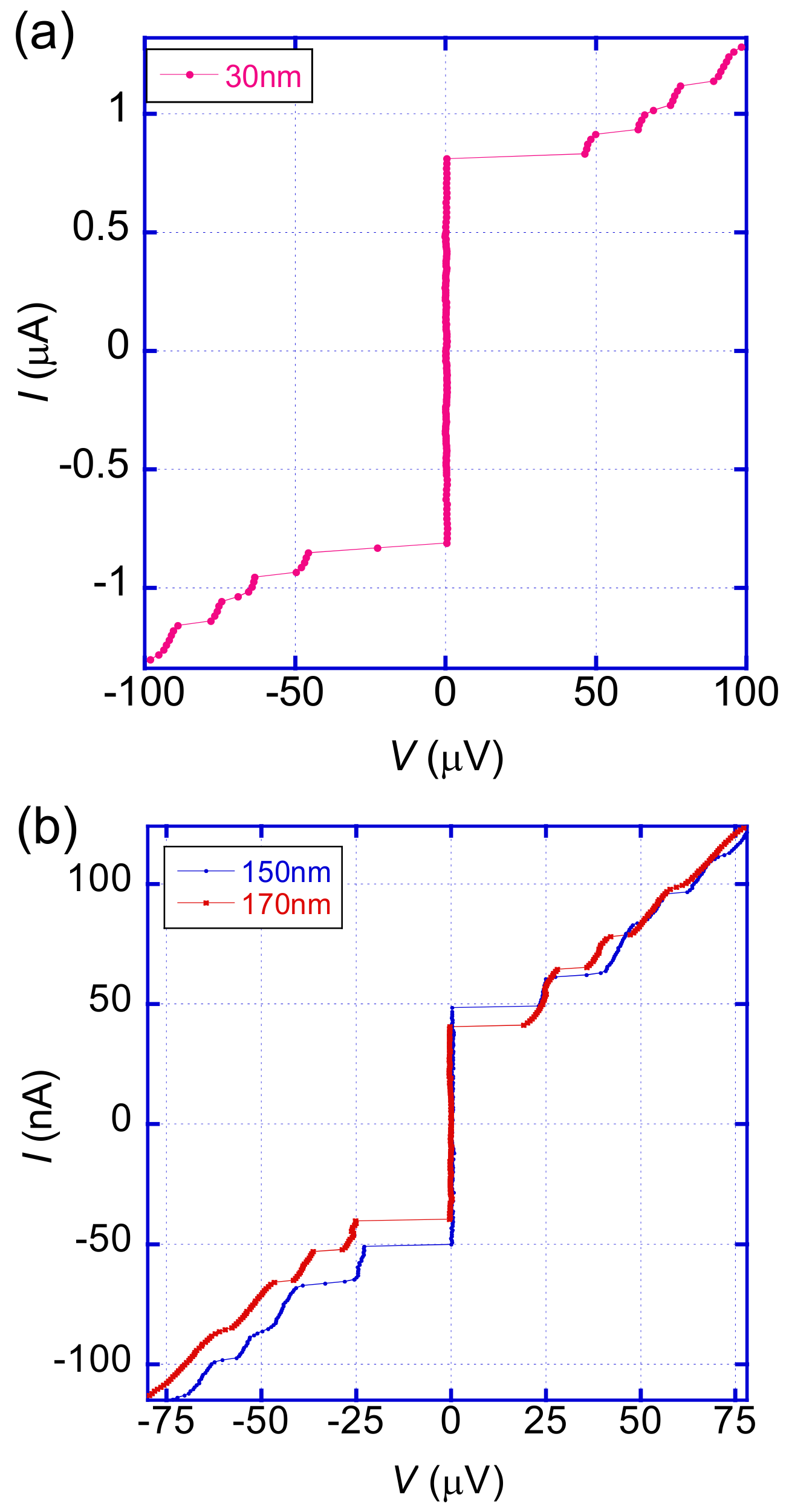}
\caption{ Current voltage characteristics at low voltages for device $A_1$ ( length  $L=30$\,nm and normal state resistance $R_{n}=0.16$\,k$\Omega$). The device
shows successive voltage jumps with the resistance continuously increasing just after each voltage jumps. b) Similarly, the current-voltage characteristics for 
somewhat longer devices $B_{5a}$ and $B_{5b}$ ($L= 150$, and 170\,nm), show similar voltage steps at low voltages.}
\label{fig:Figure8}
\end{figure}

The fact that the positions of the temperature-independent structures are the same in different junctions makes it unlikely that they are related to external
electromagnetic resonances, but rather result from some general intrinsic mechanism.
To get a better insight in the origin of  the temperature independent subgap structures we analyzed the shape of the IVC, see Fig.\,\ref{fig:Figure8}.  In all investigated
junctions the IVC have a staircase shape and consist of a number of successive voltage steps. Between the steps, the current continuously grows with the differential resistance increasing after every step. Such a behavior may be explained by successive emergence of normally conducting domains in the wire as soon as the current exceeds the critical value. This picture closely resembles the resistive states in superconducting whiskers containing phase slip 
centers (PSC).\cite{Meyer,Skocpol} 
Although one cannot in a straightforward way extend the PSC scenario in truly superconducting whiskers\cite{Galaiko} to the proximity induced 
superconductivity in nanowires, 
one cannot exclude the possibility of formation of some kind of spatially inhomogeneous resistive state in the proximity region.

\section{Gate dependence}

In this section, we investigate the gate dependence of the IVC in the superconducting state of suspended devices of type C shown in Fig. 1c and in Table I. 
The data presented in the in the previous sections are obtained at the zero gate voltage for the conduction regime with multiple open conducting channels. Here we discuss the change of the IVCs in this regime with variation of the gate voltage. An opposite, few-channel transport regime 
at large negative gate voltage, showing 
quantization of the normal conductance and the Josephson critical current was investigated in
Ref. \onlinecite{Abay}.  

In our device, the gate voltage controls the local carrier concentration in the nanowire and thereby affects the strength 
of the proximity effect. Due to a strong capacitive coupling of the gate to the wire, this variation is significant allowing us 
to observe a cross over from the SNS to SIS type regime of the current transport at low temperature. According to the
theory,\cite{Bratus1995,Ingerman} the IVC of the transparent wire should exhibit, besides a large Josephson current, a 
large excess current both in the subgap voltage region and at the large voltage. On the other hand, more resistive wires
should exhibit a small Josephson current, a suppressed subgap current, and a cross over to deficit (negative excess) current 
at large voltage. 
\begin{figure}[h]
\includegraphics[width=0.75\linewidth]{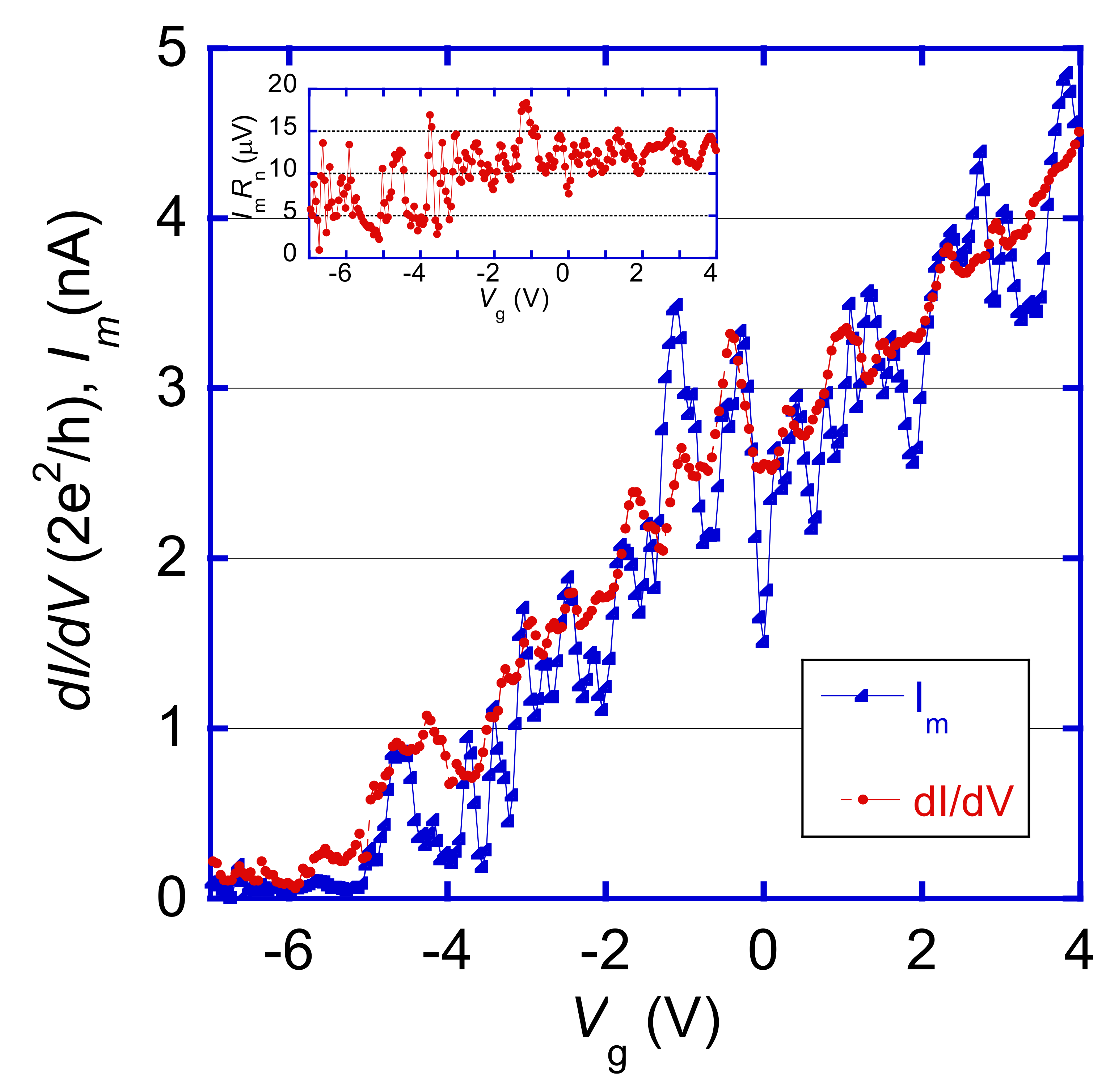}
\caption{Normal state conductance and maximum Josephson current as a function of a local-gate voltage for device 
C$_{11}$. After opening of the first conducting channel at $V_g\approx -5$V, the overall conductance and critical 
current  increase linearly with the gate voltage. In the inset the $I_mR_n$ product as a function of gate voltage. 
The constant value indicates that the maximum current is correlated with the normal state conductance.}
\label{fig:IcVg}
\end{figure} 

\begin{figure}
\includegraphics[width=0.75\linewidth]{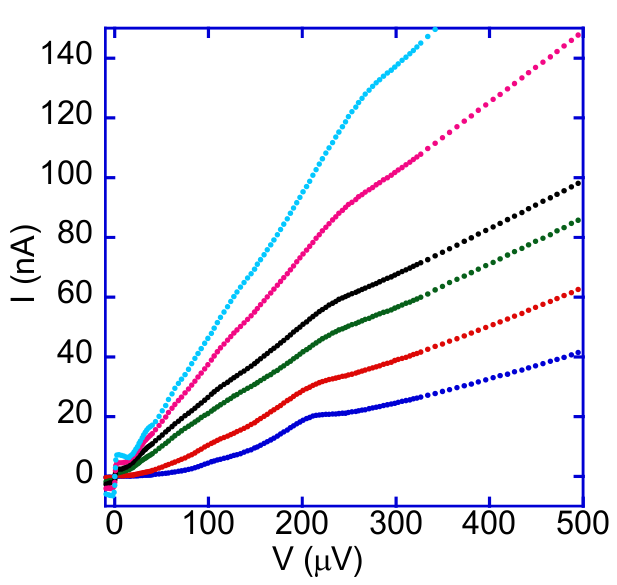}
\caption{Current-voltage characteristics for device C$_{12}$ of length $L=200$\,nm.  for different  gate voltages, 
$V_{g}=-2.16, -1.86, -1.64, -1.40, -0.93, 0.92$V (from bottom to top). The IVC exhibit cross over from the tunneling 
regime with small subgap current and negative excess current to the SNS regime with enhanced subgap conductance and positive excess current.   }
\label{fig:IVC(Vg)}
\end{figure}

The dependence of the maximum Josephson current on the gate voltage is shown in Fig.\,\ref{fig:IcVg} for the 
suspended device  $C_{11}$ ($L\approx150\,$nm). The change of the maximum current (blue) varies in tact with the 
change of the differential conductance (red). Owing to the n-type nature of the nanowires, the conductance and 
the maximum Josephson current are strongly suppressed at large negative gate voltages, $V_{g} < -5\,$V. Changing 
the gate voltage towards positive values results in linear increase of the averaged conductance and maximum Josephson 
current, the latter reaching the value of $I_m = 4$ nA at $V_g = 3\,$V. Simultaneously, the $I_mR_n$ product saturates 
at the value, $I_mR_n = 12.5\,\mu$V, and remains constant over a wide range of gate voltages, as shown in the inset in 
Fig.\,\ref{fig:IcVg}. 

In Fig.\,\ref{fig:IVC(Vg)} we present a set of IVCs for the suspended device C$_{12}$ ($L\approx$ 200\,nm and 
$R_{n}=3.6$\,k$\Omega$) for gate voltages ranging from $V_g=-2.16$V to $V_g=+0.92$V.  At large positive gate 
voltage, {\em i.e.} at large conductance, the IVC shows significant excess current and enhanced subgap conductance, 
indicating highly transmissive SNS regime. In the opposite limit of large negative gate voltage (small conductance), the 
IVC has a typical form for SIS  tunnel junctions with negative excess current\cite{Bezuglyi} and strongly suppressed
subgap conductance. The suppression of the subgap conductance is explained by small probability of MAR processes at 
small voltage, which scales with $\mathcal{D}^n$, where $n=2\Delta/eV$ is the number of Andreev reflections. In the
tunneling regime with small $\mathcal{D}\ll 1$ the subgap conductance is exponentially small. At the intermediate gate
voltages the device exhibits continuous crossover between these two regimes in accordance with the theoretical predictions
for contacts with varying transparency.\cite{Bratus1995,Ingerman}

\section{Conclusion}

We have investigated, both experimentally and theoretically, proximity effect in InAs nanowires connected to
superconducting electrodes. We have fabricated and investigated a large number of nanowire devices with suspended 
gate controlled nanowires and non-suspended nanowires, with a broad range of lengths and normal state resistances.
We measured current-voltage characteristics and analyzed their main features: the Josephson current, excess current, 
and subgap current as Äfunctions of length, temperature, magnetic field and gate voltage, and compared them with theory.
The devices show reproducible resistance per unit length, and highly transmissive interfaces.
The measured superconducting characteristics are consistent and agree reasonably well in most cases with 
theoretically computed values. The maximum Josephson current for a short length device, $L=30$\,nm, exhibits a record 
high magnitude of $800\,$nA at low temperature that comes close to the theoretically expected value.
The maximum Josephson current in other devices is typically reduced compared to the theoretical values. 
The measured excess current in most of the devices is consistent with the normal resistance data and agrees 
well with the theory. The subgap current shows large number of structures, some of them are identified as subharmonic 
gap structures generated by MAR. The other structures, detected in both suspended and non-suspended devices, have the 
form of the voltage steps at voltages that are independent of either the superconducting gap or the length of the wire. 
By varying the gate voltage in suspended devices we were able to observe a cross over from typical tunneling transport, 
with suppressed subgap current and negative excess current, at large negative gate voltage to pronounced SNS-type 
behavior, with enhanced subgap conductance and large positive excess current, at large positive gate voltage.

\begin{acknowledgments}
We acknowledge fruitful discussions with Lars Samuelson, Christopher Wilson, Thilo Bauch, and Jonas Bylander. The work was supported 
by the Swedish Research council, the Wallenberg foundation. HQX also acknowledges the National Basic Research Program 
of the Ministry of Science and Technology of China (Nos. 2012CB932703 and 2012CB932700).  
\end{acknowledgments}

%
%

\bibliography{SNS_InAsnanowire_biblio}

\begin{thebibliography}{46}%
\makeatletter
\providecommand \@ifxundefined [1]{%
 \@ifx{#1\undefined}
}%
\providecommand \@ifnum [1]{%
 \ifnum #1\expandafter \@firstoftwo
 \else \expandafter \@secondoftwo
 \fi
}%
\providecommand \@ifx [1]{%
 \ifx #1\expandafter \@firstoftwo
 \else \expandafter \@secondoftwo
 \fi
}%
\providecommand \natexlab [1]{#1}%
\providecommand \enquote  [1]{``#1''}%
\providecommand \bibnamefont  [1]{#1}%
\providecommand \bibfnamefont [1]{#1}%
\providecommand \citenamefont [1]{#1}%
\providecommand \href@noop [0]{\@secondoftwo}%
\providecommand \href [0]{\begingroup \@sanitize@url \@href}%
\providecommand \@href[1]{\@@startlink{#1}\@@href}%
\providecommand \@@href[1]{\endgroup#1\@@endlink}%
\providecommand \@sanitize@url [0]{\catcode `\\12\catcode `\$12\catcode
  `\&12\catcode `\#12\catcode `\^12\catcode `\_12\catcode `\%12\relax}%
\providecommand \@@startlink[1]{}%
\providecommand \@@endlink[0]{}%
\providecommand \url  [0]{\begingroup\@sanitize@url \@url }%
\providecommand \@url [1]{\endgroup\@href {#1}{\urlprefix }}%
\providecommand \urlprefix  [0]{URL }%
\providecommand \Eprint [0]{\href }%
\providecommand \doibase [0]{http://dx.doi.org/}%
\providecommand \selectlanguage [0]{\@gobble}%
\providecommand \bibinfo  [0]{\@secondoftwo}%
\providecommand \bibfield  [0]{\@secondoftwo}%
\providecommand \translation [1]{[#1]}%
\providecommand \BibitemOpen [0]{}%
\providecommand \bibitemStop [0]{}%
\providecommand \bibitemNoStop [0]{.\EOS\space}%
\providecommand \EOS [0]{\spacefactor3000\relax}%
\providecommand \BibitemShut  [1]{\csname bibitem#1\endcsname}%
\let\auto@bib@innerbib\@empty
\bibitem [{\citenamefont {Samuelson}(2003)}]{samuelsson}%
  \BibitemOpen
  \bibfield  {author} {\bibinfo {author} {\bibfnamefont {L.}~\bibnamefont
  {Samuelson}},\ }\href {\doibase
  http://dx.doi.org/10.1016/S1369-7021(03)01026-5} {\bibfield  {journal}
  {\bibinfo  {journal} {Mater. Today}\ }\textbf {\bibinfo {volume} {6}},\
  \bibinfo {pages} {22 } (\bibinfo {year} {2003})}\BibitemShut {NoStop}%
\bibitem [{\citenamefont {Thelander}\ \emph {et~al.}(2006)\citenamefont
  {Thelander}, \citenamefont {Agarwal}, \citenamefont {Brongersma},
  \citenamefont {Eymery}, \citenamefont {Feiner}, \citenamefont {Forchel},
  \citenamefont {Scheffler}, \citenamefont {Riess}, \citenamefont {Ohlsson},
  \citenamefont {Gösele},\ and\ \citenamefont {Samuelson}}]{clas}%
  \BibitemOpen
  \bibfield  {author} {\bibinfo {author} {\bibfnamefont {C.}~\bibnamefont
  {Thelander}}, \bibinfo {author} {\bibfnamefont {P.}~\bibnamefont {Agarwal}},
  \bibinfo {author} {\bibfnamefont {S.}~\bibnamefont {Brongersma}}, \bibinfo
  {author} {\bibfnamefont {J.}~\bibnamefont {Eymery}}, \bibinfo {author}
  {\bibfnamefont {L.}~\bibnamefont {Feiner}}, \bibinfo {author} {\bibfnamefont
  {A.}~\bibnamefont {Forchel}}, \bibinfo {author} {\bibfnamefont
  {M.}~\bibnamefont {Scheffler}}, \bibinfo {author} {\bibfnamefont
  {W.}~\bibnamefont {Riess}}, \bibinfo {author} {\bibfnamefont
  {B.}~\bibnamefont {Ohlsson}}, \bibinfo {author} {\bibfnamefont
  {U.}~\bibnamefont {Gösele}}, \ and\ \bibinfo {author} {\bibfnamefont
  {L.}~\bibnamefont {Samuelson}},\ }\href {\doibase
  http://dx.doi.org/10.1016/S1369-7021(06)71651-0} {\bibfield  {journal}
  {\bibinfo  {journal} {Mater. Today}\ }\textbf {\bibinfo {volume} {9}},\
  \bibinfo {pages} {28 } (\bibinfo {year} {2006})}\BibitemShut {NoStop}%
\bibitem [{\citenamefont {Li}\ \emph {et~al.}(2006)\citenamefont {Li},
  \citenamefont {Qian}, \citenamefont {Xiang},\ and\ \citenamefont
  {Lieber}}]{Lieber}%
  \BibitemOpen
  \bibfield  {author} {\bibinfo {author} {\bibfnamefont {Y.}~\bibnamefont
  {Li}}, \bibinfo {author} {\bibfnamefont {F.}~\bibnamefont {Qian}}, \bibinfo
  {author} {\bibfnamefont {J.}~\bibnamefont {Xiang}}, \ and\ \bibinfo {author}
  {\bibfnamefont {C.~M.}\ \bibnamefont {Lieber}},\ }\href {\doibase
  http://dx.doi.org/10.1016/S1369-7021(06)71650-9} {\bibfield  {journal}
  {\bibinfo  {journal} {Mater. Today}\ }\textbf {\bibinfo {volume} {9}},\
  \bibinfo {pages} {18 } (\bibinfo {year} {2006})}\BibitemShut {NoStop}%
\bibitem [{\citenamefont {Dayeh}(2010)}]{Shadi}%
  \BibitemOpen
  \bibfield  {author} {\bibinfo {author} {\bibfnamefont {S.~A.}\ \bibnamefont
  {Dayeh}},\ }\href@noop {} {\bibfield  {journal} {\bibinfo  {journal}
  {Semicond. Sci. and Technol.}\ }\textbf {\bibinfo {volume} {25}},\ \bibinfo
  {pages} {024004} (\bibinfo {year} {2010})}\BibitemShut {NoStop}%
\bibitem [{\citenamefont {van Weperen}\ \emph {et~al.}(2013)\citenamefont {van
  Weperen}, \citenamefont {Plissard}, \citenamefont {Bakkers}, \citenamefont
  {Frolov},\ and\ \citenamefont {Kouwenhoven}}]{Kouwenhoven}%
  \BibitemOpen
  \bibfield  {author} {\bibinfo {author} {\bibfnamefont {I.}~\bibnamefont {van
  Weperen}}, \bibinfo {author} {\bibfnamefont {S.~R.}\ \bibnamefont
  {Plissard}}, \bibinfo {author} {\bibfnamefont {E.~P. A.~M.}\ \bibnamefont
  {Bakkers}}, \bibinfo {author} {\bibfnamefont {S.~M.}\ \bibnamefont {Frolov}},
  \ and\ \bibinfo {author} {\bibfnamefont {L.~P.}\ \bibnamefont
  {Kouwenhoven}},\ }\href {\doibase 10.1021/nl3035256} {\bibfield  {journal}
  {\bibinfo  {journal} {Nano Lett.}\ }\textbf {\bibinfo {volume} {13}},\
  \bibinfo {pages} {387} (\bibinfo {year} {2013})}\BibitemShut {NoStop}%
\bibitem [{\citenamefont {Abay}\ \emph {et~al.}(2013)\citenamefont {Abay},
  \citenamefont {Persson}, \citenamefont {Nilsson}, \citenamefont {Xu},
  \citenamefont {Fogelstr\"{o}m}, \citenamefont {Shumeiko},\ and\ \citenamefont
  {Delsing}}]{simon}%
  \BibitemOpen
  \bibfield  {author} {\bibinfo {author} {\bibfnamefont {S.}~\bibnamefont
  {Abay}}, \bibinfo {author} {\bibfnamefont {D.}~\bibnamefont {Persson}},
  \bibinfo {author} {\bibfnamefont {H.}~\bibnamefont {Nilsson}}, \bibinfo
  {author} {\bibfnamefont {H.~Q.}\ \bibnamefont {Xu}}, \bibinfo {author}
  {\bibfnamefont {M.}~\bibnamefont {Fogelstr\"{o}m}}, \bibinfo {author}
  {\bibfnamefont {V.}~\bibnamefont {Shumeiko}}, \ and\ \bibinfo {author}
  {\bibfnamefont {P.}~\bibnamefont {Delsing}},\ }\href {\doibase
  10.1021/nl4014265} {\bibfield  {journal} {\bibinfo  {journal} {Nano Lett.}\
  }\textbf {\bibinfo {volume} {13}},\ \bibinfo {pages} {3614} (\bibinfo {year}
  {2013})}\BibitemShut {NoStop}%
\bibitem [{\citenamefont {Jespersen}\ \emph {et~al.}(2009)\citenamefont
  {Jespersen}, \citenamefont {Polianski}, \citenamefont {S\o{}rensen},
  \citenamefont {Flensberg},\ and\ \citenamefont {Nyg{\aa}rd}}]{Jespersen}%
  \BibitemOpen
  \bibfield  {author} {\bibinfo {author} {\bibfnamefont {T.~S.}\ \bibnamefont
  {Jespersen}}, \bibinfo {author} {\bibfnamefont {M.~L.}\ \bibnamefont
  {Polianski}}, \bibinfo {author} {\bibfnamefont {C.~B.}\ \bibnamefont
  {S\o{}rensen}}, \bibinfo {author} {\bibfnamefont {K.}~\bibnamefont
  {Flensberg}}, \ and\ \bibinfo {author} {\bibfnamefont {J.}~\bibnamefont
  {Nyg{\aa}rd}},\ }\href@noop {} {\bibfield  {journal} {\bibinfo  {journal}
  {New J.Phys.}\ }\textbf {\bibinfo {volume} {11}},\ \bibinfo {pages} {113025}
  (\bibinfo {year} {2009})}\BibitemShut {NoStop}%
\bibitem [{\citenamefont {Xiang}\ \emph {et~al.}(2006)\citenamefont {Xiang},
  \citenamefont {Vidan}, \citenamefont {Tinkham}, \citenamefont {Westervelt},\
  and\ \citenamefont {Lieber}}]{Tinkham}%
  \BibitemOpen
  \bibfield  {author} {\bibinfo {author} {\bibfnamefont {J.}~\bibnamefont
  {Xiang}}, \bibinfo {author} {\bibfnamefont {A.}~\bibnamefont {Vidan}},
  \bibinfo {author} {\bibfnamefont {M.}~\bibnamefont {Tinkham}}, \bibinfo
  {author} {\bibfnamefont {R.~M.}\ \bibnamefont {Westervelt}}, \ and\ \bibinfo
  {author} {\bibfnamefont {M.}~\bibnamefont {Lieber}, \bibfnamefont
  {Charles}},\ }\href {\doibase 10.1038/nnano.2006.140} {\bibfield  {journal}
  {\bibinfo  {journal} {Nat. Nanotechnol.}\ }\textbf {\bibinfo {volume} {1}},\
  \bibinfo {pages} {208} (\bibinfo {year} {2006})}\BibitemShut {NoStop}%
\bibitem [{\citenamefont {Doh}\ \emph {et~al.}(2005)\citenamefont {Doh},
  \citenamefont {van Dam}, \citenamefont {Roest}, \citenamefont {Bakkers},
  \citenamefont {Kouwenhoven},\ and\ \citenamefont {De~Franceschi}}]{Kewo}%
  \BibitemOpen
  \bibfield  {author} {\bibinfo {author} {\bibfnamefont {Y.-J.}\ \bibnamefont
  {Doh}}, \bibinfo {author} {\bibfnamefont {J.~A.}\ \bibnamefont {van Dam}},
  \bibinfo {author} {\bibfnamefont {A.~L.}\ \bibnamefont {Roest}}, \bibinfo
  {author} {\bibfnamefont {E.~P. A.~M.}\ \bibnamefont {Bakkers}}, \bibinfo
  {author} {\bibfnamefont {L.~P.}\ \bibnamefont {Kouwenhoven}}, \ and\ \bibinfo
  {author} {\bibfnamefont {S.}~\bibnamefont {De~Franceschi}},\ }\href {\doibase
  10.1126/science.1113523} {\bibfield  {journal} {\bibinfo  {journal}
  {Science}\ }\textbf {\bibinfo {volume} {309}},\ \bibinfo {pages} {272}
  (\bibinfo {year} {2005})}\BibitemShut {NoStop}%
\bibitem [{\citenamefont {Nilsson}\ \emph {et~al.}(2012)\citenamefont
  {Nilsson}, \citenamefont {Samuelsson}, \citenamefont {Caroff},\ and\
  \citenamefont {Xu}}]{Xu}%
  \BibitemOpen
  \bibfield  {author} {\bibinfo {author} {\bibfnamefont {H.~A.}\ \bibnamefont
  {Nilsson}}, \bibinfo {author} {\bibfnamefont {P.}~\bibnamefont {Samuelsson}},
  \bibinfo {author} {\bibfnamefont {P.}~\bibnamefont {Caroff}}, \ and\ \bibinfo
  {author} {\bibfnamefont {H.~Q.}\ \bibnamefont {Xu}},\ }\href {\doibase
  10.1021/nl203380w} {\bibfield  {journal} {\bibinfo  {journal} {Nano Lett.}\
  }\textbf {\bibinfo {volume} {12}},\ \bibinfo {pages} {228} (\bibinfo {year}
  {2012})}\BibitemShut {NoStop}%
\bibitem [{\citenamefont {Abay}\ \emph {et~al.}(2012)\citenamefont {Abay},
  \citenamefont {Nilsson}, \citenamefont {Wu}, \citenamefont {Xu},
  \citenamefont {Wilson},\ and\ \citenamefont {Delsing}}]{Abay}%
  \BibitemOpen
  \bibfield  {author} {\bibinfo {author} {\bibfnamefont {S.}~\bibnamefont
  {Abay}}, \bibinfo {author} {\bibfnamefont {H.}~\bibnamefont {Nilsson}},
  \bibinfo {author} {\bibfnamefont {F.}~\bibnamefont {Wu}}, \bibinfo {author}
  {\bibfnamefont {H.}~\bibnamefont {Xu}}, \bibinfo {author} {\bibfnamefont
  {C.}~\bibnamefont {Wilson}}, \ and\ \bibinfo {author} {\bibfnamefont
  {P.}~\bibnamefont {Delsing}},\ }\href {\doibase 10.1021/nl302740f} {\bibfield
   {journal} {\bibinfo  {journal} {Nano Lett.}\ }\textbf {\bibinfo {volume}
  {12}},\ \bibinfo {pages} {5622} (\bibinfo {year} {2012})}\BibitemShut
  {NoStop}%
\bibitem [{\citenamefont {Nishio}\ \emph {et~al.}(2011)\citenamefont {Nishio},
  \citenamefont {Kozakai}, \citenamefont {Amaha}, \citenamefont {Larsson},
  \citenamefont {Nilsson}, \citenamefont {Xu}, \citenamefont {Zhang},
  \citenamefont {Tateno}, \citenamefont {Takayanagi},\ and\ \citenamefont
  {Ishibashi}}]{Takahiro}%
  \BibitemOpen
  \bibfield  {author} {\bibinfo {author} {\bibfnamefont {T.}~\bibnamefont
  {Nishio}}, \bibinfo {author} {\bibfnamefont {T.}~\bibnamefont {Kozakai}},
  \bibinfo {author} {\bibfnamefont {S.}~\bibnamefont {Amaha}}, \bibinfo
  {author} {\bibfnamefont {M.}~\bibnamefont {Larsson}}, \bibinfo {author}
  {\bibfnamefont {H.~A.}\ \bibnamefont {Nilsson}}, \bibinfo {author}
  {\bibfnamefont {H.~Q.}\ \bibnamefont {Xu}}, \bibinfo {author} {\bibfnamefont
  {G.}~\bibnamefont {Zhang}}, \bibinfo {author} {\bibfnamefont
  {K.}~\bibnamefont {Tateno}}, \bibinfo {author} {\bibfnamefont
  {H.}~\bibnamefont {Takayanagi}}, \ and\ \bibinfo {author} {\bibfnamefont
  {K.}~\bibnamefont {Ishibashi}},\ }\href@noop {} {\bibfield  {journal}
  {\bibinfo  {journal} {Nanotechnology}\ }\textbf {\bibinfo {volume} {22}},\
  \bibinfo {pages} {445701} (\bibinfo {year} {2011})}\BibitemShut {NoStop}%
\bibitem [{\citenamefont {Doh}\ \emph {et~al.}(2008)\citenamefont {Doh},
  \citenamefont {Franceschi}, \citenamefont {Bakkers},\ and\ \citenamefont
  {Kouwenhoven}}]{Doh}%
  \BibitemOpen
  \bibfield  {author} {\bibinfo {author} {\bibfnamefont {Y.-J.}\ \bibnamefont
  {Doh}}, \bibinfo {author} {\bibfnamefont {S.~D.}\ \bibnamefont {Franceschi}},
  \bibinfo {author} {\bibfnamefont {E.~P. A.~M.}\ \bibnamefont {Bakkers}}, \
  and\ \bibinfo {author} {\bibfnamefont {L.~P.}\ \bibnamefont {Kouwenhoven}},\
  }\href {\doibase 10.1021/nl801454k} {\bibfield  {journal} {\bibinfo
  {journal} {Nano Lett.}\ }\textbf {\bibinfo {volume} {8}},\ \bibinfo {pages}
  {4098} (\bibinfo {year} {2008})}\BibitemShut {NoStop}%
\bibitem [{\citenamefont {Hofstetter}\ \emph {et~al.}(2009)\citenamefont
  {Hofstetter}, \citenamefont {Csonka}, \citenamefont {Nyg{\aa}rd},\ and\
  \citenamefont {Sch\"{o}nenberger}}]{Schoenenberger}%
  \BibitemOpen
  \bibfield  {author} {\bibinfo {author} {\bibfnamefont {L.}~\bibnamefont
  {Hofstetter}}, \bibinfo {author} {\bibfnamefont {S.}~\bibnamefont {Csonka}},
  \bibinfo {author} {\bibfnamefont {J.}~\bibnamefont {Nyg{\aa}rd}}, \ and\
  \bibinfo {author} {\bibfnamefont {C.}~\bibnamefont {Sch\"{o}nenberger}},\
  }\href {\doibase 10.1038/nature08432} {\bibfield  {journal} {\bibinfo
  {journal} {Nature}\ }\textbf {\bibinfo {volume} {461}},\ \bibinfo {pages}
  {960} (\bibinfo {year} {2009})}\BibitemShut {NoStop}%
\bibitem [{\citenamefont {Mourik}\ \emph {et~al.}(2012)\citenamefont {Mourik},
  \citenamefont {Zuo}, \citenamefont {Frolov}, \citenamefont {Plissard},
  \citenamefont {Bakkers},\ and\ \citenamefont {Kouwenhoven}}]{LeoMaj}%
  \BibitemOpen
  \bibfield  {author} {\bibinfo {author} {\bibfnamefont {V.}~\bibnamefont
  {Mourik}}, \bibinfo {author} {\bibfnamefont {K.}~\bibnamefont {Zuo}},
  \bibinfo {author} {\bibfnamefont {S.~M.}\ \bibnamefont {Frolov}}, \bibinfo
  {author} {\bibfnamefont {S.~R.}\ \bibnamefont {Plissard}}, \bibinfo {author}
  {\bibfnamefont {E.~P. A.~M.}\ \bibnamefont {Bakkers}}, \ and\ \bibinfo
  {author} {\bibfnamefont {L.~P.}\ \bibnamefont {Kouwenhoven}},\ }\href
  {\doibase 10.1126/science.1222360} {\bibfield  {journal} {\bibinfo  {journal}
  {Science}\ }\textbf {\bibinfo {volume} {336}},\ \bibinfo {pages} {1003}
  (\bibinfo {year} {2012})}\BibitemShut {NoStop}%
\bibitem [{\citenamefont {Deng}\ \emph {et~al.}(2012)\citenamefont {Deng},
  \citenamefont {Yu}, \citenamefont {Huang}, \citenamefont {Larsson},
  \citenamefont {Caroff},\ and\ \citenamefont {Xu}}]{LundMaj}%
  \BibitemOpen
  \bibfield  {author} {\bibinfo {author} {\bibfnamefont {M.~T.}\ \bibnamefont
  {Deng}}, \bibinfo {author} {\bibfnamefont {C.~L.}\ \bibnamefont {Yu}},
  \bibinfo {author} {\bibfnamefont {G.~Y.}\ \bibnamefont {Huang}}, \bibinfo
  {author} {\bibfnamefont {M.}~\bibnamefont {Larsson}}, \bibinfo {author}
  {\bibfnamefont {P.}~\bibnamefont {Caroff}}, \ and\ \bibinfo {author}
  {\bibfnamefont {H.~Q.}\ \bibnamefont {Xu}},\ }\href {\doibase
  10.1021/nl303758w} {\bibfield  {journal} {\bibinfo  {journal} {Nano Lett.}\
  }\textbf {\bibinfo {volume} {12}},\ \bibinfo {pages} {6414} (\bibinfo {year}
  {2012})}\BibitemShut {NoStop}%
\bibitem [{\citenamefont {Das}\ \emph {et~al.}(2012)\citenamefont {Das},
  \citenamefont {Ronen}, \citenamefont {Most}, \citenamefont {Oreg},
  \citenamefont {Heiblum},\ and\ \citenamefont {Shtrikman}}]{Anionic}%
  \BibitemOpen
  \bibfield  {author} {\bibinfo {author} {\bibfnamefont {A.}~\bibnamefont
  {Das}}, \bibinfo {author} {\bibfnamefont {Y.}~\bibnamefont {Ronen}}, \bibinfo
  {author} {\bibfnamefont {Y.}~\bibnamefont {Most}}, \bibinfo {author}
  {\bibfnamefont {Y.}~\bibnamefont {Oreg}}, \bibinfo {author} {\bibfnamefont
  {M.}~\bibnamefont {Heiblum}}, \ and\ \bibinfo {author} {\bibfnamefont
  {H.}~\bibnamefont {Shtrikman}},\ }\href {\doibase 10.1038/nphys2479}
  {\bibfield  {journal} {\bibinfo  {journal} {Nature Phys.}\ }\textbf {\bibinfo
  {volume} {8}},\ \bibinfo {pages} {887} (\bibinfo {year} {2012})}\BibitemShut
  {NoStop}%
\bibitem [{\citenamefont {Kulik}\ and\ \citenamefont
  {Omel'yanchuk}(1975)}]{Kulikl}%
  \BibitemOpen
  \bibfield  {author} {\bibinfo {author} {\bibfnamefont {I.}~\bibnamefont
  {Kulik}}\ and\ \bibinfo {author} {\bibfnamefont {A.}~\bibnamefont
  {Omel'yanchuk}},\ }\href@noop {} {\bibfield  {journal} {\bibinfo  {journal}
  {Sov. Phys.--JETP Letters}\ }\textbf {\bibinfo {volume} {21}},\ \bibinfo
  {pages} {96} (\bibinfo {year} {1975})}\BibitemShut {NoStop}%
\bibitem [{\citenamefont {Klapwijk}\ \emph {et~al.}(1982)\citenamefont
  {Klapwijk}, \citenamefont {Blonder},\ and\ \citenamefont
  {Tinkham}}]{Klapwijk1982}%
  \BibitemOpen
  \bibfield  {author} {\bibinfo {author} {\bibfnamefont {T.}~\bibnamefont
  {Klapwijk}}, \bibinfo {author} {\bibfnamefont {G.}~\bibnamefont {Blonder}}, \
  and\ \bibinfo {author} {\bibfnamefont {M.}~\bibnamefont {Tinkham}},\
  }\href@noop {} {\bibfield  {journal} {\bibinfo  {journal} {Physica B+C}\
  }\textbf {\bibinfo {volume} {109}},\ \bibinfo {pages} {1657} (\bibinfo {year}
  {1982})}\BibitemShut {NoStop}%
\bibitem [{\citenamefont {{Kretinin}}\ \emph {et~al.}()\citenamefont
  {{Kretinin}}, \citenamefont {{Das}},\ and\ \citenamefont
  {{Shtrikman}}}]{Andrey}%
  \BibitemOpen
  \bibfield  {author} {\bibinfo {author} {\bibfnamefont {A.}~\bibnamefont
  {{Kretinin}}}, \bibinfo {author} {\bibfnamefont {A.}~\bibnamefont {{Das}}}, \
  and\ \bibinfo {author} {\bibfnamefont {H.}~\bibnamefont {{Shtrikman}}},\
  }\href@noop {} {\ }\Eprint {http://arxiv.org/abs/1303.1410} {arXiv:1303.1410}
  \BibitemShut {NoStop}%
\bibitem [{\citenamefont {Meyer}\ and\ \citenamefont
  {Minnigerode}(1972)}]{Meyer}%
  \BibitemOpen
  \bibfield  {author} {\bibinfo {author} {\bibfnamefont {J.}~\bibnamefont
  {Meyer}}\ and\ \bibinfo {author} {\bibfnamefont {G.}~\bibnamefont
  {Minnigerode}},\ }\href {\doibase
  http://dx.doi.org/10.1016/0375-9601(72)90802-X} {\bibfield  {journal}
  {\bibinfo  {journal} {Phys. Lett. A}\ }\textbf {\bibinfo {volume} {38}},\
  \bibinfo {pages} {529 } (\bibinfo {year} {1972})}\BibitemShut {NoStop}%
\bibitem [{\citenamefont {Skocpol}\ \emph {et~al.}(1974)\citenamefont
  {Skocpol}, \citenamefont {Beasley},\ and\ \citenamefont {Tinkham}}]{Skocpol}%
  \BibitemOpen
  \bibfield  {author} {\bibinfo {author} {\bibfnamefont {W.}~\bibnamefont
  {Skocpol}}, \bibinfo {author} {\bibfnamefont {M.}~\bibnamefont {Beasley}}, \
  and\ \bibinfo {author} {\bibfnamefont {M.}~\bibnamefont {Tinkham}},\ }\href
  {\doibase 10.1007/BF00655865} {\bibfield  {journal} {\bibinfo  {journal} {J.
  Low Temp. Phys.}\ }\textbf {\bibinfo {volume} {16}},\ \bibinfo {pages} {145}
  (\bibinfo {year} {1974})}\BibitemShut {NoStop}%
\bibitem [{\citenamefont {Ohlsson}(2002)}]{LundFab}%
  \BibitemOpen
  \bibfield  {author} {\bibinfo {author} {\bibfnamefont {B.}~\bibnamefont
  {Ohlsson}},\ }\href {\doibase
  http://dx.doi.org/10.1016/S1386-9477(02)00318-1} {\bibfield  {journal}
  {\bibinfo  {journal} {Physica E}\ }\textbf {\bibinfo {volume} {13}},\
  \bibinfo {pages} {1126 } (\bibinfo {year} {2002})}\BibitemShut {NoStop}%
\bibitem [{\citenamefont {Nilsson}\ \emph {et~al.}(2008)\citenamefont
  {Nilsson}, \citenamefont {Duty}, \citenamefont {Abay}, \citenamefont
  {Wilson}, \citenamefont {Wagner}, \citenamefont {Thelander}, \citenamefont
  {Delsing},\ and\ \citenamefont {Samuelson}}]{Henrik}%
  \BibitemOpen
  \bibfield  {author} {\bibinfo {author} {\bibfnamefont {H.~A.}\ \bibnamefont
  {Nilsson}}, \bibinfo {author} {\bibfnamefont {T.}~\bibnamefont {Duty}},
  \bibinfo {author} {\bibfnamefont {S.}~\bibnamefont {Abay}}, \bibinfo {author}
  {\bibfnamefont {C.}~\bibnamefont {Wilson}}, \bibinfo {author} {\bibfnamefont
  {J.~B.}\ \bibnamefont {Wagner}}, \bibinfo {author} {\bibfnamefont
  {C.}~\bibnamefont {Thelander}}, \bibinfo {author} {\bibfnamefont
  {P.}~\bibnamefont {Delsing}}, \ and\ \bibinfo {author} {\bibfnamefont
  {L.}~\bibnamefont {Samuelson}},\ }\href {\doibase 10.1021/nl0731062}
  {\bibfield  {journal} {\bibinfo  {journal} {Nano Lett.}\ }\textbf {\bibinfo
  {volume} {8}},\ \bibinfo {pages} {872} (\bibinfo {year} {2008})}\BibitemShut
  {NoStop}%
\bibitem [{\citenamefont {Suyatin}\ \emph {et~al.}(2007)\citenamefont
  {Suyatin}, \citenamefont {Thelander}, \citenamefont {Bj\"{o}rk},
  \citenamefont {Maximov},\ and\ \citenamefont {Samuelson}}]{Lund}%
  \BibitemOpen
  \bibfield  {author} {\bibinfo {author} {\bibfnamefont {D.~B.}\ \bibnamefont
  {Suyatin}}, \bibinfo {author} {\bibfnamefont {C.}~\bibnamefont {Thelander}},
  \bibinfo {author} {\bibfnamefont {M.~T.}\ \bibnamefont {Bj\"{o}rk}}, \bibinfo
  {author} {\bibfnamefont {I.}~\bibnamefont {Maximov}}, \ and\ \bibinfo
  {author} {\bibfnamefont {L.}~\bibnamefont {Samuelson}},\ }\href@noop {}
  {\bibfield  {journal} {\bibinfo  {journal} {Nanotechnology}\ }\textbf
  {\bibinfo {volume} {18}},\ \bibinfo {pages} {105307} (\bibinfo {year}
  {2007})}\BibitemShut {NoStop}%
\bibitem [{\citenamefont {Datta}(1997)}]{datta}%
  \BibitemOpen
  \bibfield  {author} {\bibinfo {author} {\bibfnamefont {S.}~\bibnamefont
  {Datta}},\ }\href@noop {} {\emph {\bibinfo {title} {{Electronic Transport in
  Mesoscopic Systems}}}}\ (\bibinfo  {publisher} {Cambridge University Press},\
  \bibinfo {year} {1997})\BibitemShut {NoStop}%
\bibitem [{\citenamefont {Samuelsson}\ \emph {et~al.}(2004)\citenamefont
  {Samuelsson}, \citenamefont {Ingerman}, \citenamefont {Johansson},
  \citenamefont {Bezuglyi}, \citenamefont {Shumeiko}, \citenamefont {Wendin},
  \citenamefont {K\"ursten}, \citenamefont {Richter}, \citenamefont
  {Matsuyama},\ and\ \citenamefont {Merkt}}]{Peter}%
  \BibitemOpen
  \bibfield  {author} {\bibinfo {author} {\bibfnamefont {P.}~\bibnamefont
  {Samuelsson}}, \bibinfo {author} {\bibfnamefont {A.}~\bibnamefont
  {Ingerman}}, \bibinfo {author} {\bibfnamefont {G.}~\bibnamefont {Johansson}},
  \bibinfo {author} {\bibfnamefont {E.~V.}\ \bibnamefont {Bezuglyi}}, \bibinfo
  {author} {\bibfnamefont {V.~S.}\ \bibnamefont {Shumeiko}}, \bibinfo {author}
  {\bibfnamefont {G.}~\bibnamefont {Wendin}}, \bibinfo {author} {\bibfnamefont
  {R.}~\bibnamefont {K\"ursten}}, \bibinfo {author} {\bibfnamefont
  {A.}~\bibnamefont {Richter}}, \bibinfo {author} {\bibfnamefont
  {T.}~\bibnamefont {Matsuyama}}, \ and\ \bibinfo {author} {\bibfnamefont
  {U.}~\bibnamefont {Merkt}},\ }\href {\doibase 10.1103/PhysRevB.70.212505}
  {\bibfield  {journal} {\bibinfo  {journal} {Phys. Rev. B}\ }\textbf {\bibinfo
  {volume} {70}},\ \bibinfo {pages} {212505} (\bibinfo {year}
  {2004})}\BibitemShut {NoStop}%
\bibitem [{\citenamefont {Furusaki}\ \emph {et~al.}(1992)\citenamefont
  {Furusaki}, \citenamefont {Takayanagi},\ and\ \citenamefont
  {Tsukada}}]{Furusaki}%
  \BibitemOpen
  \bibfield  {author} {\bibinfo {author} {\bibfnamefont {A.}~\bibnamefont
  {Furusaki}}, \bibinfo {author} {\bibfnamefont {H.}~\bibnamefont
  {Takayanagi}}, \ and\ \bibinfo {author} {\bibfnamefont {M.}~\bibnamefont
  {Tsukada}},\ }\href {\doibase 10.1103/PhysRevB.45.10563} {\bibfield
  {journal} {\bibinfo  {journal} {Phys. Rev. B}\ }\textbf {\bibinfo {volume}
  {45}},\ \bibinfo {pages} {10563} (\bibinfo {year} {1992})}\BibitemShut
  {NoStop}%
\bibitem [{\citenamefont {Arnold}(1987)}]{Arnold1987}%
  \BibitemOpen
  \bibfield  {author} {\bibinfo {author} {\bibfnamefont {G.}~\bibnamefont
  {Arnold}},\ }\href {\doibase 10.1007/BF00682620} {\bibfield  {journal}
  {\bibinfo  {journal} {J. Low Temp. Phys.}\ }\textbf {\bibinfo {volume}
  {68}},\ \bibinfo {pages} {1} (\bibinfo {year} {1987})}\BibitemShut {NoStop}%
\bibitem [{\citenamefont {Bratus'}\ \emph {et~al.}(1995)\citenamefont
  {Bratus'}, \citenamefont {Shumeiko},\ and\ \citenamefont
  {Wendin}}]{Bratus1995}%
  \BibitemOpen
  \bibfield  {author} {\bibinfo {author} {\bibfnamefont {E.~N.}\ \bibnamefont
  {Bratus'}}, \bibinfo {author} {\bibfnamefont {V.~S.}\ \bibnamefont
  {Shumeiko}}, \ and\ \bibinfo {author} {\bibfnamefont {G.}~\bibnamefont
  {Wendin}},\ }\href {\doibase 10.1103/PhysRevLett.74.2110} {\bibfield
  {journal} {\bibinfo  {journal} {Phys. Rev. Lett.}\ }\textbf {\bibinfo
  {volume} {74}},\ \bibinfo {pages} {2110} (\bibinfo {year}
  {1995})}\BibitemShut {NoStop}%
\bibitem [{\citenamefont {Averin}\ and\ \citenamefont
  {Bardas}(1995)}]{Averin1995}%
  \BibitemOpen
  \bibfield  {author} {\bibinfo {author} {\bibfnamefont {D.}~\bibnamefont
  {Averin}}\ and\ \bibinfo {author} {\bibfnamefont {A.}~\bibnamefont
  {Bardas}},\ }\href {\doibase 10.1103/PhysRevLett.75.1831} {\bibfield
  {journal} {\bibinfo  {journal} {Phys. Rev. Lett.}\ }\textbf {\bibinfo
  {volume} {75}},\ \bibinfo {pages} {1831} (\bibinfo {year}
  {1995})}\BibitemShut {NoStop}%
\bibitem [{\citenamefont {Cuevas}\ \emph {et~al.}(1996)\citenamefont {Cuevas},
  \citenamefont {Mart\'{i}n-Rodero},\ and\ \citenamefont
  {Yeyati}}]{Cuevas1996}%
  \BibitemOpen
  \bibfield  {author} {\bibinfo {author} {\bibfnamefont {J.~C.}\ \bibnamefont
  {Cuevas}}, \bibinfo {author} {\bibfnamefont {A.}~\bibnamefont
  {Mart\'{i}n-Rodero}}, \ and\ \bibinfo {author} {\bibfnamefont {A.~L.}\
  \bibnamefont {Yeyati}},\ }\href {\doibase 10.1103/PhysRevB.54.7366}
  {\bibfield  {journal} {\bibinfo  {journal} {Phys. Rev. B}\ }\textbf {\bibinfo
  {volume} {54}},\ \bibinfo {pages} {7366} (\bibinfo {year}
  {1996})}\BibitemShut {NoStop}%
\bibitem [{com()}]{comment}%
  \BibitemOpen
  \href@noop {} {}\bibinfo {note} {Cross over from diffusive short- to
  long-junction regime could alternatively be modeled with a theory developed
  in Ref. \onlinecite{Bezuglyi}.}\BibitemShut {Stop}%
\bibitem [{\citenamefont {Cuevas}\ and\ \citenamefont
  {Fogelstr\"om}(2001)}]{Cuevas2001}%
  \BibitemOpen
  \bibfield  {author} {\bibinfo {author} {\bibfnamefont {J.~C.}\ \bibnamefont
  {Cuevas}}\ and\ \bibinfo {author} {\bibfnamefont {M.}~\bibnamefont
  {Fogelstr\"om}},\ }\href {\doibase 10.1103/PhysRevB.64.104502} {\bibfield
  {journal} {\bibinfo  {journal} {Phys. Rev. B}\ }\textbf {\bibinfo {volume}
  {64}},\ \bibinfo {pages} {104502} (\bibinfo {year} {2001})}\BibitemShut
  {NoStop}%
\bibitem [{\citenamefont {Eschrig}(2009)}]{Eschrig2009}%
  \BibitemOpen
  \bibfield  {author} {\bibinfo {author} {\bibfnamefont {M.}~\bibnamefont
  {Eschrig}},\ }\href {\doibase 10.1103/PhysRevB.80.134511} {\bibfield
  {journal} {\bibinfo  {journal} {Phys. Rev. B}\ }\textbf {\bibinfo {volume}
  {80}},\ \bibinfo {pages} {134511} (\bibinfo {year} {2009})}\BibitemShut
  {NoStop}%
\bibitem [{\citenamefont {Zaitsev}(1984)}]{Zaitsev1984}%
  \BibitemOpen
  \bibfield  {author} {\bibinfo {author} {\bibfnamefont {A.~V.}\ \bibnamefont
  {Zaitsev}},\ }\href@noop {} {\bibfield  {journal} {\bibinfo  {journal} {Sov.
  Phys. --JETP}\ }\textbf {\bibinfo {volume} {59}},\ \bibinfo {pages} {1015}
  (\bibinfo {year} {1984})}\BibitemShut {NoStop}%
\bibitem [{\citenamefont {Shumeiko}\ \emph {et~al.}(1997)\citenamefont
  {Shumeiko}, \citenamefont {Bratus},\ and\ \citenamefont
  {Wendin}}]{Shumeiko1997}%
  \BibitemOpen
  \bibfield  {author} {\bibinfo {author} {\bibfnamefont {V.~S.}\ \bibnamefont
  {Shumeiko}}, \bibinfo {author} {\bibfnamefont {E.~N.}\ \bibnamefont
  {Bratus}}, \ and\ \bibinfo {author} {\bibfnamefont {G.}~\bibnamefont
  {Wendin}},\ }\href {\doibase http://dx.doi.org/10.1063/1.593475} {\bibfield
  {journal} {\bibinfo  {journal} {Low Temp. Phys.}\ }\textbf {\bibinfo {volume}
  {23}},\ \bibinfo {pages} {181} (\bibinfo {year} {1997})}\BibitemShut
  {NoStop}%
\bibitem [{\citenamefont {Cuevas}\ \emph {et~al.}(2006)\citenamefont {Cuevas},
  \citenamefont {Hammer}, \citenamefont {Kopu}, \citenamefont {Viljas},\ and\
  \citenamefont {Eschrig}}]{Cuevas}%
  \BibitemOpen
  \bibfield  {author} {\bibinfo {author} {\bibfnamefont {J.~C.}\ \bibnamefont
  {Cuevas}}, \bibinfo {author} {\bibfnamefont {J.}~\bibnamefont {Hammer}},
  \bibinfo {author} {\bibfnamefont {J.}~\bibnamefont {Kopu}}, \bibinfo {author}
  {\bibfnamefont {J.~K.}\ \bibnamefont {Viljas}}, \ and\ \bibinfo {author}
  {\bibfnamefont {M.}~\bibnamefont {Eschrig}},\ }\href {\doibase
  10.1103/PhysRevB.73.184505} {\bibfield  {journal} {\bibinfo  {journal} {Phys.
  Rev. B}\ }\textbf {\bibinfo {volume} {73}},\ \bibinfo {pages} {184505}
  (\bibinfo {year} {2006})}\BibitemShut {NoStop}%
\bibitem [{\citenamefont {Bezuglyi}\ \emph {et~al.}(2011)\citenamefont
  {Bezuglyi}, \citenamefont {Bratus'},\ and\ \citenamefont
  {Shumeiko}}]{Bezuglyi}%
  \BibitemOpen
  \bibfield  {author} {\bibinfo {author} {\bibfnamefont {E.~V.}\ \bibnamefont
  {Bezuglyi}}, \bibinfo {author} {\bibfnamefont {E.~N.}\ \bibnamefont
  {Bratus'}}, \ and\ \bibinfo {author} {\bibfnamefont {V.~S.}\ \bibnamefont
  {Shumeiko}},\ }\href {\doibase 10.1103/PhysRevB.83.184517} {\bibfield
  {journal} {\bibinfo  {journal} {Phys. Rev. B}\ }\textbf {\bibinfo {volume}
  {83}},\ \bibinfo {pages} {184517} (\bibinfo {year} {2011})}\BibitemShut
  {NoStop}%
\bibitem [{\citenamefont {Martinis}\ and\ \citenamefont
  {Grabert}(1988)}]{Martinis1988}%
  \BibitemOpen
  \bibfield  {author} {\bibinfo {author} {\bibfnamefont {J.~M.}\ \bibnamefont
  {Martinis}}\ and\ \bibinfo {author} {\bibfnamefont {H.}~\bibnamefont
  {Grabert}},\ }\href {\doibase 10.1103/PhysRevB.38.2371} {\bibfield  {journal}
  {\bibinfo  {journal} {Phys. Rev. B}\ }\textbf {\bibinfo {volume} {38}},\
  \bibinfo {pages} {2371} (\bibinfo {year} {1988})}\BibitemShut {NoStop}%
\bibitem [{\citenamefont {Ambegaokar}\ and\ \citenamefont
  {Halperin}(1969)}]{Ambegaokar1969}%
  \BibitemOpen
  \bibfield  {author} {\bibinfo {author} {\bibfnamefont {V.}~\bibnamefont
  {Ambegaokar}}\ and\ \bibinfo {author} {\bibfnamefont {B.~I.}\ \bibnamefont
  {Halperin}},\ }\href {\doibase 10.1103/PhysRevLett.22.1364} {\bibfield
  {journal} {\bibinfo  {journal} {Phys. Rev. Lett.}\ }\textbf {\bibinfo
  {volume} {22}},\ \bibinfo {pages} {1364} (\bibinfo {year}
  {1969})}\BibitemShut {NoStop}%
\bibitem [{\citenamefont {Bladh}\ \emph {et~al.}(2003)\citenamefont {Bladh},
  \citenamefont {Gunnarsson}, \citenamefont {H{\"u}rfeld}, \citenamefont
  {Devi}, \citenamefont {Kristoffersson}, \citenamefont {Sm{\aa}lander},
  \citenamefont {Pehrson}, \citenamefont {Claeson}, \citenamefont {Delsing},\
  and\ \citenamefont {Taslakov}}]{Bladh2003}%
  \BibitemOpen
  \bibfield  {author} {\bibinfo {author} {\bibfnamefont {K.}~\bibnamefont
  {Bladh}}, \bibinfo {author} {\bibfnamefont {D.}~\bibnamefont {Gunnarsson}},
  \bibinfo {author} {\bibfnamefont {E.}~\bibnamefont {H{\"u}rfeld}}, \bibinfo
  {author} {\bibfnamefont {S.}~\bibnamefont {Devi}}, \bibinfo {author}
  {\bibfnamefont {C.}~\bibnamefont {Kristoffersson}}, \bibinfo {author}
  {\bibfnamefont {B.}~\bibnamefont {Sm{\aa}lander}}, \bibinfo {author}
  {\bibfnamefont {S.}~\bibnamefont {Pehrson}}, \bibinfo {author} {\bibfnamefont
  {T.}~\bibnamefont {Claeson}}, \bibinfo {author} {\bibfnamefont
  {P.}~\bibnamefont {Delsing}}, \ and\ \bibinfo {author} {\bibfnamefont
  {M.}~\bibnamefont {Taslakov}},\ }\href@noop {} {\bibfield  {journal}
  {\bibinfo  {journal} {Review of Scientific Instruments}\ }\textbf {\bibinfo
  {volume} {74}},\ \bibinfo {pages} {1323} (\bibinfo {year}
  {2003})}\BibitemShut {NoStop}%
\bibitem [{\citenamefont {Kupriyanov}\ \emph {et~al.}(1999)\citenamefont
  {Kupriyanov}, \citenamefont {Brinkman}, \citenamefont {Golubov},
  \citenamefont {Siegel},\ and\ \citenamefont {Rogalla}}]{Kupriyanov1999}%
  \BibitemOpen
  \bibfield  {author} {\bibinfo {author} {\bibfnamefont {M.~Y.}\ \bibnamefont
  {Kupriyanov}}, \bibinfo {author} {\bibfnamefont {A.}~\bibnamefont
  {Brinkman}}, \bibinfo {author} {\bibfnamefont {A.~A.}\ \bibnamefont
  {Golubov}}, \bibinfo {author} {\bibfnamefont {M.}~\bibnamefont {Siegel}}, \
  and\ \bibinfo {author} {\bibfnamefont {H.}~\bibnamefont {Rogalla}},\
  }\href@noop {} {\bibfield  {journal} {\bibinfo  {journal} {Physica C}\
  }\textbf {\bibinfo {volume} {326}},\ \bibinfo {pages} {16} (\bibinfo {year}
  {1999})}\BibitemShut {NoStop}%
\bibitem [{\citenamefont {Blonder}\ \emph {et~al.}(1982)\citenamefont
  {Blonder}, \citenamefont {Tinkham},\ and\ \citenamefont
  {Klapwijk}}]{Blonder1982}%
  \BibitemOpen
  \bibfield  {author} {\bibinfo {author} {\bibfnamefont {G.~E.}\ \bibnamefont
  {Blonder}}, \bibinfo {author} {\bibfnamefont {M.}~\bibnamefont {Tinkham}}, \
  and\ \bibinfo {author} {\bibfnamefont {T.~M.}\ \bibnamefont {Klapwijk}},\
  }\href {\doibase 10.1103/PhysRevB.25.4515} {\bibfield  {journal} {\bibinfo
  {journal} {Phys. Rev. B}\ }\textbf {\bibinfo {volume} {25}},\ \bibinfo
  {pages} {4515} (\bibinfo {year} {1982})}\BibitemShut {NoStop}%
\bibitem [{\citenamefont {Galaiko}\ and\ \citenamefont
  {Kopnin}(1986)}]{Galaiko}%
  \BibitemOpen
  \bibfield  {author} {\bibinfo {author} {\bibfnamefont {V.}~\bibnamefont
  {Galaiko}}\ and\ \bibinfo {author} {\bibfnamefont {N.}~\bibnamefont
  {Kopnin}},\ }in\ \href@noop {} {\emph {\bibinfo {booktitle} {Nonequilibrium
  Superconductivity}}},\ \bibinfo {editor} {edited by\ \bibinfo {editor}
  {\bibfnamefont {D.}~\bibnamefont {Langenberg}}\ and\ \bibinfo {editor}
  {\bibfnamefont {A.}~\bibnamefont {Larkin}}}\ (\bibinfo  {publisher}
  {North-Holland},\ \bibinfo {address} {Amsterdam},\ \bibinfo {year}
  {1986})\BibitemShut {NoStop}%
\bibitem [{\citenamefont {Ingerman}\ \emph {et~al.}(2001)\citenamefont
  {Ingerman}, \citenamefont {Johansson}, \citenamefont {Shumeiko},\ and\
  \citenamefont {Wendin}}]{Ingerman}%
  \BibitemOpen
  \bibfield  {author} {\bibinfo {author} {\bibfnamefont {A.}~\bibnamefont
  {Ingerman}}, \bibinfo {author} {\bibfnamefont {G.}~\bibnamefont {Johansson}},
  \bibinfo {author} {\bibfnamefont {V.~S.}\ \bibnamefont {Shumeiko}}, \ and\
  \bibinfo {author} {\bibfnamefont {G.}~\bibnamefont {Wendin}},\ }\href
  {\doibase 10.1103/PhysRevB.64.144504} {\bibfield  {journal} {\bibinfo
  {journal} {Phys. Rev. B}\ }\textbf {\bibinfo {volume} {64}},\ \bibinfo
  {pages} {144504} (\bibinfo {year} {2001})}\BibitemShut {NoStop}%
\end{thebibliography}%

\end{document}